# Integrated Cooling (i-Cool) Textile of Heat Conduction and Sweat Transportation for Personal Perspiration Management

Yucan Peng[1†], Wei Li[2†], Bofei Liu[1], Joseph Schaadt[3,4], Jing Tang[1], Guangmin Zhou[1], Weiliang Jin[2], Yangying Zhu[1], Guanyang Wang[5], Wenxiao Huang[1], Chi Zhang[6], Tong Wu[1], Chris Dames[3,4], Ravi Prasher[3,4], Shanhui Fan[2] & Yi Cui[1,7*]

[1]Department of Materials Science and Engineering, Stanford University, Stanford, CA 94305, USA.

[2]E. L. Ginzton Laboratory, Department of Electrical Engineering, Stanford University, Stanford, CA 94305, USA.

[3]Department of Mechanical Engineering, University of California, Berkeley, CA 94720, USA

[4] Energy Technologies Area, Lawrence Berkeley National Laboratory, 1 Cyclotron Road, Berkeley, CA 94720, USA

[5]Department of Mathematics, Stanford University, Stanford, CA 94305, USA

[6]Department of Mechanical Engineering, Stanford University, Stanford, CA 94305, USA.

[7]Stanford Institute for Materials and Energy Sciences, SLAC National Accelerator Laboratory, 2575 Sand Hill Road, Menlo Park, CA 94025, USA.

[†]These authors contributed equally to this work

***Corresponding author: Yi Cui (yicui@stanford.edu)**

**Abstract**

**Perspiration evaporation plays an indispensable role in human body heat dissipation. However, conventional textiles show limited perspiration management capability in moderate/profuse perspiration scenarios, i.e. low evaporation ability, ineffective evaporative cooling effect, and resultant human body dehydration and electrolyte disorder. Here, we propose a novel concept of integrated cooling (i-Cool) textile of heat conduction and sweat transportation for personal perspiration management based on unique functional structure design. By integrating heat conductive pathways and water transport channels decently, this textile not only shows the capability of liquid water wicking, but also exhibits superior evaporation rate than traditional textiles. Furthermore, compared with cotton, about 2.8 °C cooling effect causing less than one third amount of dehydration has also been demonstrated on the artificial sweating skin platform with feedback control loop simulating human body perspiration situation. Moreover, the practical application feasibility of the i-Cool textile design principles has been validated as well. Owing to its exceptional personal perspiration management performance in liquid water wicking, fast evaporation, efficient cooling effect and reduced human body dehydration/electrolyte loss, we expect this i-Cool textile provides promising design guidelines for next-generation personal perspiration management textile.**

Satisfaction with the thermal environment for human body is significant, not merely due to the demand for comfort, but more importantly because thermal conditions are crucial for human body health[1,2]. Heat-resulted physiological and psychological problems not only can be threatening for human health[3], but also negatively influence labor productivity and society economy[4]. Personal thermal management focusing on thermal conditions of human body and its local environment is emerging as an energy-efficient and cost-effective solution[5-7]. Without consuming excess energy on managing the temperature of the entire environment[8,9], innovative textile design and development in controlling human body heat dissipation routes is one of the most effective and direct paths of personal thermal management[10-12]. In general, human body dissipates heat via four different pathways: radiation, convection, conduction and evaporation[13]. Recently, textiles with engineered radiative properties[14-19], convective and conductive properties[20-22] have been demonstrated as promising approaches for personal thermal management[23]. However, textiles for effective personal perspiration or evaporation management is still lacking.

For the delicate human body system with a narrow temperature range (36 – 38 °C core temperature at rest and up to 41°C for heavy exercise)[24], evaporation plays an indispensable role in human body thermoregulation. Even at a very mild scenario, about 20 percent of heat dissipation of the dry human body relies on the water vapour loss via insensible perspiration[13,25]. With further increase of heat load, liquid sweat evaporation on the skin contributes to more and more heat loss and becomes the major route for human body heat dissipation in intense scenarios such as heavy exercise and hot/humid environments, where excess heat cannot be dissipated efficiently by other pathways[26,27]. The extremely high latent heat of water vaporization (~2440 KJ/Kg at 298 K) is the main reason why sweat evaporation can efficiently cool down human body[28]. However, if desired cooling effect is not achieved, the human body will continue generating more sweat for evaporation. Excessive perspiration can result in human body dehydration and electrolyte disorder, which is harmful and even life-threatening[29]. Besides, when people are in highly active scenarios, the maximum cooling power of sweat evaporation that can be achieved limits the maximum activity level. Therefore, to efficiently release the evaporative cooling power of sweat for the human body can not only avoid meaningless excessive perspiration, but also expand the human body activity/adaptation limit.

Nevertheless, the performance of conventional textiles is sub-optimal in moderate/profuse perspiration situations in which liquid sweat is present on the skin (See Supplementary Note 1 and Extended Data Fig. 1-2 for more discussion on dry and slight perspiration scenarios). Cotton, and moisture management fabrics, usually focus on expediting removal of sweat from human body skin with their natural or engineered strong liquid water wicking ability[30,31]. The efficient removal of sweat alleviates human body discomfort of wet and sticky sense with the buffer effect of textile, and enlarges the evaporation area for sweat. However, sweat within the conventional textiles cannot be efficiently evaporated by the human body heat due to the low thermal conductance. Also, while sweat evaporation does happen on the textiles, due to the low thermal conductance, only textile surface rather than human skin can be efficiently cooled. Hence, the cooling power associated with evaporation cannot be efficiently delivered to the human body. In other words, the sweat absorbed by the conventional textiles becomes less effective for cooling down the human body[25]. The inefficient cooling effect will lead to further perspiration, and meanwhile the slow sweat evaporation will result in the accumulation of sweat in the textile. This process will undermine the buffer effect of the textiles once the absorption limit of the fabric is reached, at which point the human body will get wet and sticky again. The further perspiration can result in potential risk of dehydration, electrolyte disorder, physical and mental deterioration or even death[32]. Therefore, a textile for personal perspiration management which is capable of fast wicking, rapid sweat evaporation, efficient cooling of skin and reducing body water loss is essential for personal perspiration management.

**i-Cool functional structure design**

In order to realize an ideal textile for personal perspiration management, we propose a novel concept of integrated cooling (i-Cool) textile of heat conduction and sweat transportation, as illustrated in Fig. 1a. We introduce heat conductive components into the textile and divide the functionalities of heat conduction and sweat transport into two operational components. The heat conductive matrix and sweat transportation channels are integrated together in the i-Cool textile. The synergistic effect of the two components result in excellent performance at sweat wicking, fast evaporation, evaporative cooling and reducing human body dehydration. As shown in Fig. 1b, the sweat transport channels can pull liquid water up from skin and spreading it on the top surface for evaporation. On the other hand, the heat conductive matrix transfers heat from skin to the top evaporation layer very efficiently. Accordingly, combined with large evaporation area and efficient heat conduction from skin, sweat wicked onto the top surface can be evaporated quickly into air, taking away a huge amount of heat from the skin. Moreover, due to the great heat conduction capability of the heat conductive matrix, the evaporative cooling effect can in return decrease skin temperature, which will consequently reduce human body dehydration. As illustrated in Fig. 1c, compared to the normal textiles, the i-Cool textile functions not only to absorb sweat but also provide an excellent heat conduction path for the evaporation cooling to take away a great amount of heat from the skin. Furthermore, the enhanced evaporation ability and efficient cooling effect can prevent the i-Cool textile from flooding to a much greater extent and avoid excessive perspiration. The improved evaporative cooling effect does not mean more sweat needs to be generated. Therefore, the i-Cool textile can help human body achieve enhanced cooling effect, by greatly reduced sweat consumed and by using the sweat in a highly efficient manner.

## Results and discussion

On the basis of the i-Cool functional structure design principles as outlined above, we selected copper (Cu) and nylon 6 nanofibres for proof of concept. It is worthwhile to mention that Cu and nylon 6 nanofibres are not the only choices. Other materials satisfying the design principles can be applied as well. Here, Cu is well-known for its extraordinary thermal conductivity ($\sim 400$ W·m$^{-1}$·K$^{-1}$), and nylon 6 nanofibres are capable of water wicking. As illustrated in Extended Data Fig. 3, electrospinning was utilized to generate nylon 6 nanofibres, which were transferred to the heat conductive Cu matrix prepared by laser cutting. With press lamination, the i-Cool (Cu) textile with desired functional structure design was fabricated. The photograph of as-fabricated i-Cool (Cu) textile is displayed in Fig. 2a. Nylon 6 nanofibres not only cover the Cu top surface, but also fill inside the pores, as shown in the magnified photograph of the bottom side of the i-Cool (Cu) textile in the inset of Fig. 2b. As the results of the press lamination process, nanofibres on the skeleton of Cu matrix are denser with smaller space among the nanofibres than the ones in the pores of Cu matrix, which can be clearly observed in the scanning electron microscope (SEM) images in Fig. 2b and Extended Data Fig. 4. The capillarity difference resulted from the morphology difference can potentially benefit evaporation since liquid water transport to the nanofibres right on the heat conductive Cu matrix is preferential[33]. To evaluate the performance of the i-Cool (Cu) textile, we selected cotton textile as the main control textile since it is arguably the most widely used textile in human history. We have also chosen other well-known activewear fabrics for comparison purposes.

**Wicking performance**

Textiles designed for perspiration scenarios must be able to transport sweat quickly from the skin (in contact with textile bottom) to textile top surface. Correspondingly, we tested in parallel the i-Cool (Cu) textile and commercial textiles including cotton, Dri-FIT, CoolMax and Coolswitch via mimicking the sweat transport process from the human body skin to the outer surface of the textile. Textile samples covered a certain amount of liquid water on the platform respectively, and the wicking rate was calculated via dividing wicking area by wicking time for every sample (Extended Data Fig. 5). It turned out that the interconnected nylon 6 nanofibres in the i-Cool (Cu) textile was able to quickly transport liquid water from bottom to top and spread it out, which exhibited comparable or higher wicking rate in comparison with conventional textiles (Fig. 2c).

**Heat transport performance**

To quantify the enhancement of heat transport capability of the i-Cool (Cu) textile, we performed the measurement of thermal resistance using cut bar method, as illustrated in Extended Data Fig. 6. Using this method, we measured the thermal resistance of the i-Cool (Cu) textile and other commercial textile samples all under a contact pressure of ~15 psi (103 kPa). As exhibited in Fig. 2d, the i-Cool (Cu) textile shows about 14 – 20 times lower thermal resistance compared to the conventional textiles due to the Cu heat transport matrix (See Supplementary Note 2 and Extended Data Fig. 6-7 for more details and discussion).

**Evaporation performance**

With demonstrated wicking performance and heat transport ability, we further used a straightforward evaporation test to compare the evaporation performance of our designed i-Cool (Cu) textile and the conventional textiles. Figure 2e illustrates the experimental setup: A heater placed on an insulating foam was used to simulate human skin with a thermocouple attached to the heater surface; We added liquid water at 37 °C to mimic sweat onto the artificial skin, then textile samples covered on the wet artificial skin immediately; During the whole evaporation process, skin temperature was always monitored and recorded. For example, a group of typical curves of skin temperature versus time are shown in Extended Data Fig. 8. Generally, the curves can be divided into three stages for the tested textile samples. Initially, when water was just added onto the artificial skin, skin temperature dropped sharply. Then, skin temperature was relatively stable only fluctuating in a small range in the evaporation stage. Eventually, skin temperature rose again quickly once water was completely evaporated.

Two pieces of important information can be obtained through comparing the curves of i-Cool (Cu) textile and the conventional textiles. Firstly, the evaporation time with i-Cool (Cu) textile was much shorter, which indicates that i-Cool (Cu) textile exhibits higher evaporation rate. This conclusion can also be verified by measuring the mass loss of liquid water over time during the evaporation test (Extended Data Fig. 9). Secondly, skin temperature with i-Cool (Cu) textile was lower than the conventional textiles during evaporation, demonstrating human body can evaporate sweat faster with even lower skin temperature when a person wears i-Cool textile. The summarized comparison of average skin temperature and average evaporation rate between the i-Cool (Cu) textile and the conventional textiles is displayed in Fig. 2f. The i-Cool (Cu) textile shows 2.3-4.5 °C lower average skin temperature and about twice faster average evaporation rate compared to the conventional textiles.

Furthermore, systematic evaporation experiments for i-Cool (Cu) textile and cotton were performed under assorted skin power density and initial liquid water amount. With different experimental parameters, the average evaporation rate was calculated and plotted versus the average skin temperature during evaporation in Extended Data Fig. 10a and Extended Data Fig. 10b. In our measurement range, a linear relationship between the average evaporation rate and the average skin temperature was observed with a certain amount of initial water. The i-Cool (Cu) textile showed overall higher evaporation rate than cotton with the same initial water amount and same skin temperature due to the superior i-Cool textile functional structure design. Employed the linear fitting relationship between average evaporation rate and average skin temperature under conditions of different initial water amount and replotted from Extended Data Fig. 10, Fig. 2g shows the relationship between the average evaporation rate and the initial water amount at different skin temperatures for the i-Cool (Cu) textile and cotton. It is obvious that the i-Cool (Cu) textile shows a much higher average evaporation rate even with a low initial water amount. Both textiles tend to reach saturated evaporation rates as the initial water amount is increased to some amount.

**Artificial sweating skin platform with feedback control loop**

With the above measurements, we have demonstrated the enhanced evaporation performance of i-Cool (Cu) textile with the functional structure design. Human body is capable of adjusting itself to maintain homeostasis in the means of feedback control loops[34]. Taking perspiration as an example, when the human body temperature exceeds a threshold, the sympathetic nervous system stimulates the eccrine sweat glands to secrete water to the skin surface. In reverse, water evaporation on the skin surface accelerates heat loss and thus body temperature decreases, which will reduce or suspend the perspiration of human body (Fig. 3a)[35,36].

To characterize the cooling effect of the i-Cool (Cu) textile and the conventional textiles, we designed an artificial sweating skin platform with feedback control loop to perform measurements simulating human body perspiration situation, as shown in Fig. 3b. In this system, an artificial sweating skin that can generate sweat uniformly from every fabricated perspiration spot was built up and served as the test platform. Power was supplied to the artificial sweating skin platform to generate heat flux simulating human body metabolic heat. A syringe pump and a temperature controller were utilized to provide continuous liquid water supply at a constant temperature (37 °C) for the artificial sweating skin. A thermocouple was attached to the artificial sweating skin platform surface, monitoring skin temperature with a thermocouple meter that transmitted skin temperature data to the computer in real time. Subsequently, the internal set program could instantly alternate the pumping rate of the syringe pump that corresponds to the sweating rate of artificial sweating skin, which realized the feedback control loop imitating human body's feedback control mechanism.

To achieve uniform water outflow through each artificial sweat pore mimicking human body skin sweating, we designed the artificial sweating skin platform as illustrated in Fig. 3c. In the bottom, an enclosed cuboid cavity connecting to water inlet acted as a water reservoir. When water was pumped in, water in the reservoir was forced out upwards through the channels on the reservoir cap. On the top of it, a perforated hydrophilic heater was attached to generate heat, in the meantime through which water can flow out. However, such kind of design cannot satisfy the requirement of uniform sweating (See Supplementary Note 3 and Extended Data Fig. 11 for more details).

Therefore, we fabricated a Janus-type wicking layer with limited water outlets and added it onto the perforated heater layer as the skin surface to realize uniform perspiration from each artificial sweat pore. To manufacture it, a mask was placed on the wicking layer, then diluted polydimethylsiloxane (PDMS) solution was sprayed on the masked wicking layer. After removing the mask, drying and curing, the uncovered top surface of the wicking layer was modified to be hydrophobic (Extended Data Fig. 12). As illustrated in the red dash box in Fig. 3c, water can diffuse into the unmodified bottom layer with strong wicking ability and be transported to the top surface, while the hydrophobic "baffles" on the top surface will confine water outflow to the unmodified hydrophilic locations. Accordingly, water wicked from the bottom can flow out only from the limited water outlets uniformly to mimic human body perspiration situation (Extended Data Fig. 13).

We believe that the measurement results obtained with the as-built artificial sweating skin platform can provide reasonable parallel cooling effect comparison among the textile samples, even though this set-up cannot fully represent the human body due to the lack of some other feedback control mechanisms such as blood flow feedback control and the differences in size, shape, thermal capacity, etc. With the realization of scale-up, we expect to conduct the human physiological wear experiment[37] in the near future.

**Artificial sweating skin test**

With the facility of artificial sweating skin platform combined with feedback control loop, we then performed measurements for the i-Cool (Cu) textile and the conventional textiles. On the artificial sweating skin platform, we first performed a demonstration experiment to show the sweat utilization efficiency difference. In this experiment, the same power density was used for the i-Cool (Cu) textile and cotton textile while the sweating rate was varied for different ones to realize the same skin temperature (34.5 °C), then we observed the condition of the artificial skin device and the textile samples after stabilization of 30 minutes. As shown in Extended Data Fig. 14, bare skin remained almost dry. The skin with the i-Cool (Cu) textile also remained dry while there was a little water absorbed in the sample. Nevertheless, there was a much larger amount of water remaining on both the skin platform and the cotton textile. These results intuitively demonstrated the i-Cool (Cu) textile can cool down the skin more efficiently consuming much less sweat.

Figure 3d shows the experimental results obtained in the measurement scenario, in which segmented power density was applied to skin heater to mimic the different metabolic rates of a human body before/after exercise and during exercise. The skin temperature and the sweating rate with different textiles were recorded spanned the whole measurement process. The skin temperature with the i-Cool (Cu) textile is lower than the skin temperature with cotton showing ~2.8 °C temperature discrepancy (~ 2 °C temperature difference with Dri-FIT and Coolswitch, ~3.4 °C temperature difference with CoolMax). The skin temperature with the i-Cool (Cu) textile is very close to bare skin. These results demonstrate the great cooling effect of the i-Cool (Cu) textile and validates the superiority of the i-Cool functional structure design during perspiration. Apart from the superior cooling effect, the sweating rate with the i-Cool (Cu) textile is also much lower than that with other textiles. Taking the cotton textile as an example, the sweat amount with i-Cool textile (Cu) is less than one third of that with cotton textile during the one-hour measurement period via integrating the sweating rate over the time span, which proves that human body dehydration can be alleviated with the i-Cool (Cu) textile. Also, we summarized the skin temperature data of the stable stage at the higher power density in multiple-time measurements. The data distribution histograms are exhibited in Extended Data Fig. 15. Extended Data Figure 15a shows the data distribution of various samples, which clearly demonstrates experimental reliability and the significant difference of the i-Cool (Cu) textile and the conventional textiles. Data distribution histogram of multiple tests for the i-Cool (Cu) textile is displayed in Extended Data Fig. 15b, indicating the good repeatability of the measurements.

Furthermore, we developed a thermal simulation based on the actual human body (See Supplementary Note 4 and Extended Data Fig. 16 for more details)[38]. The simulation results show that the i-Cool (Cu) textile can achieve temperature reduction in both the skin temperature and core temperature of the human body compared to that with cotton textile (Extended Data Fig. 16b), which further validates the potential of the i-Cool structure design in efficient evaporative cooling for the human body.

Different cotton samples with various area mass density were also tested (See Supplementary Note 5 and Extended Data Fig. 17 for more details). In our experiments, the lightest cotton sample (26.5 g/m$^2$) that is too transparent to be practically used still exhibits around 1.5 °C higher skin temperature than the i-Cool (Cu) textile. Plus, we tested the Cu heat conductive matrix and nylon 6 nanofibre film separately. The departure of the heat conduction component and water transport component makes both of them less efficient in evaporative cooling, as exhibited in Extended Data Fig. 18. These tests illustrate the key factor to achieve an effective cooling effect is the integrated functional design of heat conduction and sweat transportation.

Thus far these results have demonstrated that the i-Cool (Cu) textile can offer an exceptional cooling effect and meanwhile reduce water loss of human body during perspiration. Even though less sweat is utilized by the i-Cool (Cu) textile for evaporative cooling, it still exhibits improved cooling effect than other textiles, which implies that i-Cool structure design is capable of utilizing perspiration much more efficiently to achieve higher cooling power.

The measurement results with varied power density of the artificial sweating skin to mimic various exercise conditions are shown in Fig. 3e. The i-Cool (Cu) textile demonstrates a better cooling effect and causes less water loss under each condition of different exercise intensities, which demonstrates the i-Cool textile functional structure design is suitable for a wide range of exercise conditions where personal perspiration management is desired. Comparing the skin temperature with the i-Cool (Cu) textile and cotton under different power densities, the i-Cool (Cu) textile can achieve the same skin temperature as cotton withstanding ~150 W/m$^2$ higher power density than cotton in the artificial sweating skin test.

Besides, the evaluation of performance under diverse ambient environment conditions was performed, especially in high temperature circumstances and high relative humidity surroundings in which perspiration is more likely to happen. At the ambient temperature of 40 °C, the evaporative cooling performance of i-Cool (Cu) textile and the conventional textiles is shown in Fig. 3f. The skin temperature distinction between the i-Cool (Cu) textile and the conventional textiles is still very evident. To take a step further, we decreased skin power density of the artificial sweating skin to make skin temperature lower than ambient temperature comparing bare skin, i-Cool (Cu) and cotton, to see if the high thermal conductivity design in the i-Cool (Cu) textile will cause adverse effect for skin temperature. As a result, skin temperature with the i-Cool (Cu) textile is almost the same and shows better performance than cotton, as shown in Extended Data Fig. 19, indicating its evaporative cooling effect surpassed the opposing heat conduction from ambient.

In addition to high ambient temperature, we also investigated the performance of i-Cool (Cu) textile and other conventional textiles in a varying relative humidity environment. The experimental process is illustrated in Extended Data Fig. 20, in which a humidifier and a chamber were employed to increase ambient relative humidity. Generally, as the relative humidity was raised, skin temperature with all the textile swatches rose correspondingly (Fig. 3g). Nevertheless, the skin temperature of the i-Cool (Cu) textile was much lower than the conventional textiles.

Moreover, we performed measurements to see how the parameters in the functional structure design of i-Cool (Cu) textile influence its performance (See Supplementary Note 6 and Extended Data Fig. 21 for more details). The results provide additional guidelines for personal perspiration management textile design.

**i-Cool practical application demonstration**

To demonstrate the feasibility of applying our i-Cool textile functional structure design to practical usage in daily life, we fabricated the i-Cool textile based on a commercial fabric. For this purpose, we replaced the Cu heat conductive matrix by metal coated polyester (PET) matrix and nanoporous polyethylene (NanoPE) matrix both of which have been utilized as textile materials. Here, silver (Ag) electroless plating was used to offer the metal coating. As shown in Extended Data Fig. 22, these alternatives of Cu matrix can also maintain almost the same performance, indicating the viability of applying heat conductive coating on normal textiles to serve as the heat conductive matrix. Therefore, we further demonstrated the i-Cool textile based on a commercial fabric made of PET, which is a widely used textile material. As illustrated in Fig. 4a, a commercial PET fabric (Extended Data Fig. 23a) acting as the original fabric went through a facile electroless plating process, after which Ag coating was deposited onto every fibre's surface of the PET fabric (Extended Data Fig. 23b)[15, 39, 40]. Next, prepared nylon 6 nanofibre film was transferred onto the Ag coated PET fabric via press lamination to realize the i-Cool (Ag) textile which possessed the desired i-Cool structure. It is worthwhile to point out the PET fabric we selected and the electroless plating method are not the only choices. Other textile material and other methods offering heat conductive coatings can be utilized. Alternatively, heat conductive fibres can be applied as well for the heat transport matrix. Figure 4b shows the photograph of the i-Cool (Ag) textile sample swatch, and its optical microscope image viewing from the bottom is exhibited in Fig. 4c. The SEM image of nylon 6 nanofibres is shown in the inset. The Ag coating (thickness ~ 10 µm) on the fabric is uniform and conformal covering the surface of each PET fibre, and nylon 6 nanofibres are loaded into the pores and on top of the metal coated fabric. The morphology difference between nanofibres in the pores and on the heat conductive matrix skeleton can also be seen in Extended Data Fig. 24.

Successively, we performed the same measurement on the artificial sweating skin platform with feedback control loop for the i-Cool (Ag) textile sample, together with the i-Cool (Cu) textile and cotton swatches. Shown in Fig. 4d, the i-Cool (Ag) textile and i-Cool (Cu) textile present comparable cooling performance for personal perspiration management. The sample swatches of the original PET fabric, Ag coated PET fabric and PET fabric-Nylon 6 were also tested as control samples shown in Extended Data Fig. 25. Due to the lack of the i-Cool structure, they do not show similar cooling effect to the i-Cool (Ag) textile.

Additionally, we performed a washing test for the i-Cool (Ag) textile to check if its structure can remain after vigorous washing processes. After 50 hours of washing, the sample swatch stayed intact (Extended Data Fig. 26). The change of total mass of the washed swatch almost cannot be detected by a balance with 0.0001g accuracy. Plus, the total Ag loss from the Ag coated PET fabric for 50-hour wash was assessed via testing washing water sample with the help of inductively coupled plasma mass spectrometry (ICP-MS), showing that only less than 0.5% by mass of Ag could be lost.

## Summary

In summary, we report a novel concept of i-Cool textile with unique functional structure design for personal perspiration management. The innovative employment of integrated water transport and heat conductive functional components together not only ensures its wicking ability, but also the fast evaporation ability, enhanced evaporative cooling effect and reduction of dehydration for human body. An artificial sweating skin platform with feedback control loop simulating human body perspiration situation was realized, on which the i-Cool (Cu) textile shows apparent cooling effect and much less water loss compared to the conventional textiles. Also, the structure advantage maintains under various conditions of exercise and ambient environment. The practical application feasibility of the i-Cool design principles is also demonstrated, exhibiting decent performance and wearability. Therefore, we expect the i-Cool textile will open a new door and even revolutionize the textile for personal perspiration management in the near future.

## Methods

**Textile preparation.** The Cu matrix used in the i-Cool (Cu) textile sample (main text) was prepared with Cu foil (~ 25 µm thickness, Pred Materials) laser cut via DPSS UV laser cutter. A pore array (2 mm diameter, 3 mm pitch) on the Cu foil was created to realize the Cu matrix. The PET matrix (~ 50 µm thickness) and NanoPE matrix (~ 25 µm thickness) were prepared by laser cutting in the same way. Nylon 6 nanofibre film was prepared by electrospinning. The nylon 6 solution system used in this work is 20 wt% nylon-6 (Sigma-Aldrich) in formic acid (Alfa Aesar). The polymer solution was loaded in a 5 mL syringe with a 22-gauge needle tip, which is connected to a voltage supply (ES30P-5W, Gamma High Voltage Research). The solution was pumped out of the needle tip using a syringe pump (Aladdin). The nanofibres were collected by a grounded copper foil (Pred Materials). The applied potential was 15 kV. The pumping rate was 0.1 mL/h. The distance between the needle tip and the collector is 20 cm. After collecting nylon 6 nanofibres of desired mass, the nylon 6 nanofibre film (~ 4.5 g/m$^2$, ~ 25 µm thickness) was transferred and laminated on the Cu matrix. A hydraulic press (MTI) was used to press nylon 6 nanofibres both into the holes and on the top of the Cu matrix. The fabricated i-Cool (Cu) was ~ 45 µm thick and 107.7 g/m$^2$. The varied parameters of the i-Cool (Cu) textile are shown in Extended Data Fig. 21. To fabricate the i-Cool (Ag) textile sample, we selected a commercial PET fabric (warp knit, single jersey, ~ 20 g/m², ~130 µm thickness, purchased from fabric.com). It was cleaned firstly and modified with polydopamine (PDA) coating for 2 h in an aqueous solution that consists of 2 g/L dopamine hydrochloride (Sigma Aldrich) and 10 mM Tris-buffer solution (pH 8.5, Teknova)[41]. For electroless plating of silver (Ag), the PDA-coated PET fabric was then dipped into a 25 g/L AgNO$_3$ solution (99.9%, Alfa Aesar) for 30 min to form the Ag seed layer. After rinsing with deionized (DI) water, the PET fabric was immersed into the plating bath solution containing 4.2 g L$^{-1}$ Ag(NH$_3$)$_2$$^+$ (made by adding 28% NH$_3$·H$_2$O dropwise into 5 g L$^{-1}$ AgNO$_3$ until the solution became clear again) and 5 g L$^{-1}$ glucose (anhydrous, EMD Millipore Chemicals)[42] for 2 hours. Next, the PET fabric was turned over and placed into a new plating bath for another 2 hours. The Ag coating process was the same for the PET matrix and NanoPE matrix. After that, nylon 6 nanofibre film was added onto it by the same process described above. The cotton textile sample was from a common short-sleeve T-shirt (100% cotton, single jersey knit, 130 g/m², ~400 µm thickness, Dockers). The Dri-FIT textile sample was from a regular Dri-FIT T-shirt (100% PET, single jersey knit, 143 g/m², ~ 400 µm thickness, Nike). The CoolMax textile sample was from a T-shirt made of 100% CoolMax Extreme polyester fibers (100% PET, single jersey knit, 166 g/m², ~ 445 µm thickness, purchased from Galls.com). The Coolswitch textile sample was from a Coolswitch T-shirt (91%PET/9% Elastane, French terry knit, 140 g/m², ~350 µm thickness, Under Armour).

**Material characterization.** The optical microscope images were taken with an Olympus optical microscope. The SEM images were taken by a FEI XL30 Sirion SEM (5 kV).

**Wicking rate measurement.** The wicking rate measurement method was based on AATCC 198 with modification. 5 cm × 5 cm textile samples were prepared ahead. 0.1 mL of distilled water was placed on the simulated skin platform by pipette. Then textile samples were covered on the water,

and the time of water reaching the circle of 1.5 cm in radius on the top surface of textile was recorded. Wicking rate was calculated using wicking area divided by wicking time.

**Thermal resistance measurement.** The cut bar method adapted from ASTM 5470 was used to measure thermal resistance. In this setup, eight thermocouples are inserted into the center of two 1 inch × 1 inch copper reference bars to measure the temperature profiles along the top and bottom bar. A resistance heater generates a heat flux which flows through the top bar followed by the sample and then the bottom bar after which the heat is dissipated into a large heat sink. The entire apparatus (top bar, sample, bottom bar) is wrapped in thermal insulation. A modest pressure of approximately 15 psi was applied at the top bar to reduce contact resistance, and no thermal grease was used due to the material porosity. The temperature profiles of the top and bottom copper bars are then used to determine both the heat flux and the temperature drop across the sample stack, which can derive the total thermal resistance ($R_{tot}$). Plotting the $R_{tot}$ versus the number of sample layers, the sample thermal resistance with contact thermal resistance between samples can be obtained from the slope of the line.

**Water vapour transmission property tests.** The upright cup testing procedure was based on ASTM E96 with modification. Medium bottles (100ml; Fisher Scientific) were filled with 80ml of distilled water, and sealed with the textile samples using open-top caps and silicone gaskets (Corning). The exposed area of the textile was 3 cm in diameter. The sealed bottles were placed into an environmental chamber in which the temperature was held at 35°C and relative humidity was 30% ± 5%. The mass of the bottles and the samples was measured periodically. By dividing the reduced mass of the water by the exposed area of the bottle (3cm in diameter), the water vapour transmission was calculated. The evaporative resistance measurement was based on ISO 11092/ASTM 1868 with modification. A heater was used to generate stable heat flux mimicking

the skin. A metal foam soaked with water was placed on the heater. A waterproof but vapour permeable film was covered on the top of the metal foam to protect the textile sample from contact with water. The whole device was thermally guarded. For different textile samples, we adjusted the heat flux to maintain the same skin temperature (35 °C) for all measurements. The ambient temperature was controlled by the water recirculation system at 35 °C, and the relative humidity was within 24 ± 4%. The evaporative resistance was calculated by $R_{ef} = \frac{(P_s - P_a) \cdot A}{H} - R_{ebp}$, where $P_s$ is the water vapour pressure at the plate surface, which can be assumed as the saturation at the temperature of the surface, $P_a$ is the water vapour pressure in the air, $A$ is the area of the plate test section, $H$ is the power input, and $R_{ebp}$ is the value measured without any textile samples.

**Water vapour thermal measurement.** The artificial sweating skin platform was utilized in this measurement. A steady power density (580 W/m$^2$) and water flow rate (0.25mL/h) were adopted. An acrylic frame (thickness: 1.5mm) with a crossing was laser cut and placed on the platform to support the textile samples avoiding the liquid water contact. Stable skin temperature was read. The ambient was 22 °C ± 0.2 °C, 40% ± 5% relative humidity.

**Evaporation test.** The skin was simulated by a polyimide insulated flexible heater (McMaster-Carr, 25 cm$^2$) which was connected to a power supply (Keithley 2400). A ribbon type hot junction thermocouple (~ 0.1 mm in diameter, K-type, Omega) was in contact with the top surface of the simulated skin to measure the skin temperature. The heater was set on a 10 cm-thick foam for heat insulation. During the tests, water (37 °C) was added onto the simulated skin and textile samples were covered on the simulated skin immediately. The skin temperatures with wet textile samples during water evaporation were measured with an assorted combination of initial water amount and generated area power density of simulated skin. The average evaporation rate was calculated by dividing the initial water amount by evaporation time. The end point of the evaporation was

defined as the inflection point between the relatively stable range and the rapid increase stage of temperature. The average skin temperature referred to the average temperature reading spanned the evaporation stage in which skin temperature was relatively stable. The mass of wet textile samples was measured by a digital balance (U. S. Solid, 0.001g accuracy) to track the water mass loss during the evaporation. The tests were all performed in an environment of 22 °C ± 0.2 °C, 40% ± 5% relative humidity.

**Fabrication of Janus-type wicking layer with limited water outlets.** A filter paper (Qualitative, Whatman) was used as the wicking layer. An acrylic board was laser cut into a mask with Epilog Fusion M2 Laser and placed on the top of the filter paper. Polydimethylsiloxane (PDMS) base and curing agent (Sylgard 184, Dow Corning) with mass ratio 10: 1 were dispersed into hexane (Fisher Scientific) with volume ratio 1: 10. The PDMS solution was sprayed onto the masked filter paper that was on a heating plate, which helped with faster volatilization of hexane. After drying and curing, the PDMS formed hydrophobic coating layer only on the uncovered place of the top surface of the filter paper, which could absorb and transport water from the bottom surface but provide limited water outlets on the top surface.

**Artificial sweating skin test with feedback control loop.** The water reservoir (5 cm × 5 cm × 2.5 mm) with water inlet (whole part size: 8 cm × 8 cm × 3.5 mm) was made by 3D printing (FlashForge Creator Pro). A cover with a 9 × 9 hole (diameter: 3 mm) array (hole array area: 5 cm × 5 cm, whole part size: 8 cm × 8 cm × 1.5 mm) was also 3D printed and bound with the water reservoir part. The water reservoir was connected to a syringe pump (Harvard, PHD 2000). The pumped water was heated at 37 °C by a heater (Omega) and a proportional–integral–derivative (PID) temperature controller (Omega). A polyimide insulated flexible heater (McMaster-Carr, 25 cm$^2$) with laser cut water outlets was adhered to the holey cover. The heater was connected to a

power supply (Keithley 2400). Then, the fabricated Janus-type wicking layer with limited water outlets was attached to the heater layer to serve as the skin surface. A ribbon type hot junction thermocouple (~ 0.1 mm in diameter, K-type, Omega) connected to a thermocouple meter (Omega) was in contact with the top surface of the Janus-type wicking layer to measure the skin temperature. The thermocouple meter, syringe pump and power supply were all controlled by a LabView program, which can alter the pumping rate (sweating rate) according to the thermometer reading (skin temperature) in real time. Before the test, the artificial sweating skin platform was filled with water in advance. The perspiration threshold skin temperature was set to be 34.5 °C, over which the sweating rate was linearly dependent on skin temperature[35,36]. The relationship between pumping rate and skin temperature was set as pumping rate (mL/h) = 0.32*skin temperature (°C) - 11.04, which was decided according to previous research and reasonable human body perspiration rate range. The whole set-up was in a space without forced convection. No chamber with cover for the set-up was used to avoid water vapour accumulation except the varied humidity test. In the varied humidity test, a humidifier was placed next to the testing platform and they are enclosed together to change the humidity. The initial air temperature in the chamber was 22 °C but about 1-2 °C reading variation of the ambient temperature thermometer was observed, perhaps due to the water vapour condensation, but no obvious influence on the skin temperature was observed. In other cases, if no ambient temperature and relative humidity are specified, the ambient temperature was 22 °C ± 0.2 °C and ambient relative humidity was 40% ± 5%.

**Washing test.** The i-Cool (Ag) textile sample (5 cm × 5 cm) was washed in clean water (80 mL) under stirring at the speed of 350 rpm for 50 h. The water before and after wash was collected (5 wt% nitric acid added) and then tested using inductively coupled plasma mass spectrometry (ICP-MS) to quantify the amount of Ag released from the textile sample during washing.

## Data availability

The data that support the findings of this study are available from the corresponding authors upon request.

**Acknowledgements**

The authors acknowledge the great help of P. Zhu, C. Lau, G. Gerboni, Z. Yu and Y. Zheng. Part of this work was performed at the Stanford Nano Shared Facilities and the Stanford Nanofabrication Facility. J.S., C.D., and R.P. acknowledge the support of the Laboratory Directed Research and Development Program (LDRD) at Lawrence Berkeley National Laboratory under contract # DE-AC02-05CH11231.


**Author contributions**

Y. C. and Y. P. conceived the idea. Y. P. designed and conducted the experiments. Y. P., W. L. and B. L. performed the feedback control loop construction and programming. W. L. and W. J. conducted the simulation. B. L. drew the schematics. J. T. and G. Z. helped with sample preparation. J. S. performed the thermal resistance measurement. G. W. helped with statistical analysis. T. W. performed ICP-MS measurement. Y. Z and C. Z. helped with laser cutting process.



**Competing financial interests**


The authors declare no competing financial interests.


**Materials & Correspondence**


Correspondence and requests for materials should be addressed to Y.C (yicui@stanford.edu).


# Figures

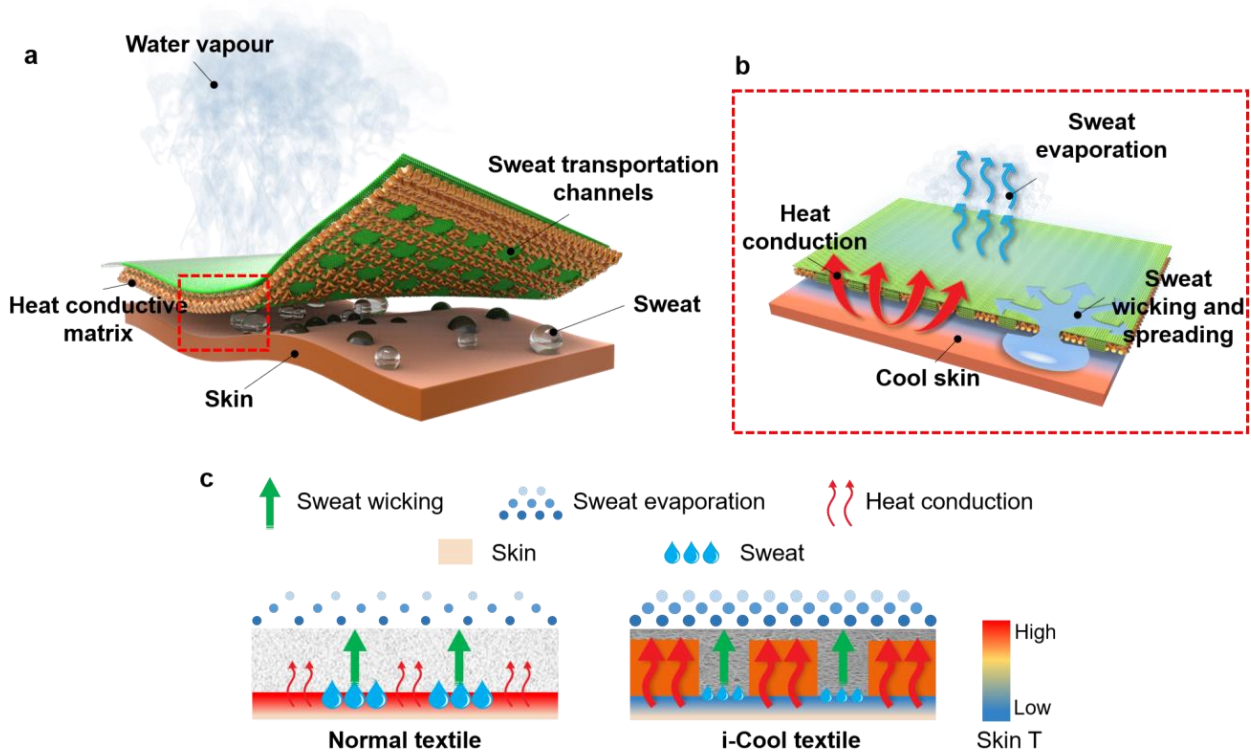

**Figure 1 | Schematic of the functional structure design of integrated cooling (i-Cool) textile of heat conduction and sweat transportation for personal perspiration management and its working mechanism. a,** Schematic of the i-Cool textile. The synergistic effect of the heat conductive matrix and sweat transport channels provides a solution to textile in personal perspiration management. **b,** Schematic of the working mechanism of the i-Cool textile. When human body perspires, the water transport channels can wick sweat from the skin surface and spread sweat onto the large-area top surface quickly. The heat conductive matrix transfers human body heat efficiently to where the evaporation happens, to assist fast evaporation. The evaporation on the i-Cool textile can deliver the evaporative cooling effect to human body skin efficiently. **c,** Comparison between normal textile and i-Cool textile. Normal textiles usually offer comfort via buffer effect of absorbing sweat, which is helpful to relieve discomfort of wet and sticky sense. However, its low thermal conductance and limited evaporation ability cannot provide effective cooling effect for skin and will undermine the buffer effect very soon. Different from normal textiles, the i-Cool textile functions not only to absorb sweat but also provide an excellent heat conduction path for the evaporation cooling to take away a great amount of heat from the skin, which can prevent the i-Cool textile from flooding to a much greater extent and avoid excessive perspiration. Therefore, the i-Cool textile can help human body

achieve enhanced cooling effect, by greatly reduced sweat consumed and by using the sweat in a highly efficient manner. The weight contrast in red arrows drawing illustrates the heat transport ability difference. The dot size and density contrast in the sweat evaporation drawing shows the different evaporation ability. The drop size contrast in the sweat drawing illustrates that i-Cool textile can help reduce sweat consumption.

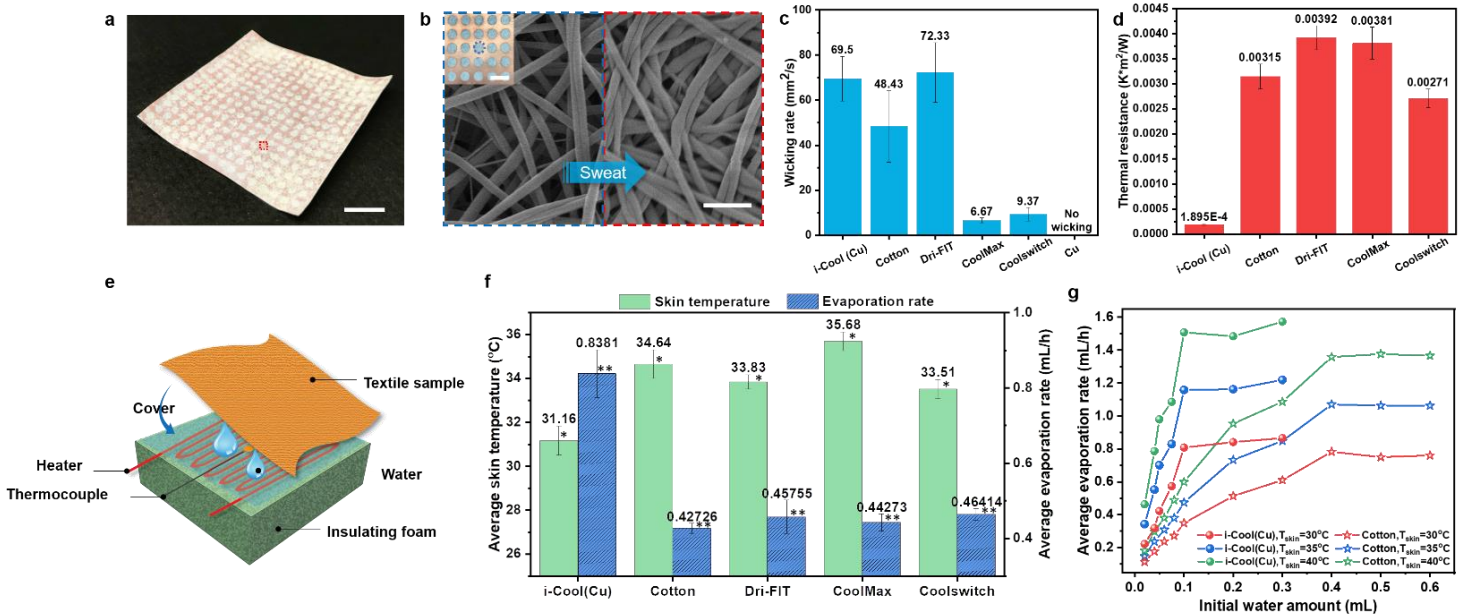

**Figure 2 | Wicking performance, thermal resistance and evaporation performance of the i-Cool (Cu) textile. a,** Photograph of as-prepared i-Cool (Cu) textile. Scale bar, 1 cm. **b,** SEM image of nylon 6 nanofibres in the pores of heat conductive matrix (blue dash box) and on the top of heat conductive matrix skeleton (red dash box). Sweat tends to be transported to the nanofibres on the heat conductive matrix skeleton due to the morphology difference. Scale bar, 1 μm. Inset is the magnified photograph of the bottom side of i-Cool (Cu) textile showing its integrated heat conduction channels and water transport channels. The holes are 2 mm in diameter and 3 mm pitch. Scale bar, 4 mm. **c,** Wicking rate of the i-Cool (Cu) textile, cotton and other commercial textiles. It shows how fast water underneath the textile can be pulled up and spread on the top surface. **d,** Thermal resistance of the i-Cool (Cu) textile (~ 45 μm thickness), cotton and other commercial textiles (~350-450 μm uncompressed thickness prior to measurement) measured by cut-bar method (See more discussion in Supplementary Note 2). **e,** Schematic of simulated skin with water for evaporation performance measurement. **f,** Average skin temperature and average evaporation rate of

the i-Cool (Cu) textile and the conventional textiles (initial water amount: 0.1 mL, skin heater power density: 422.6W/m$^2$). * Difference of average skin temperature between the i-Cool (Cu) textile and other textile samples, Welch's t-test P < 0.001, indicating strong statistical significance. ** Difference of average evaporation rate between the i-Cool (Cu) textile and other textile samples, Welch's test P < 0.001, indicating strong statistical significance. **g,** Fitted average evaporation rate of i-Cool (Cu) textile and cotton versus initial water amount at different skin temperature.

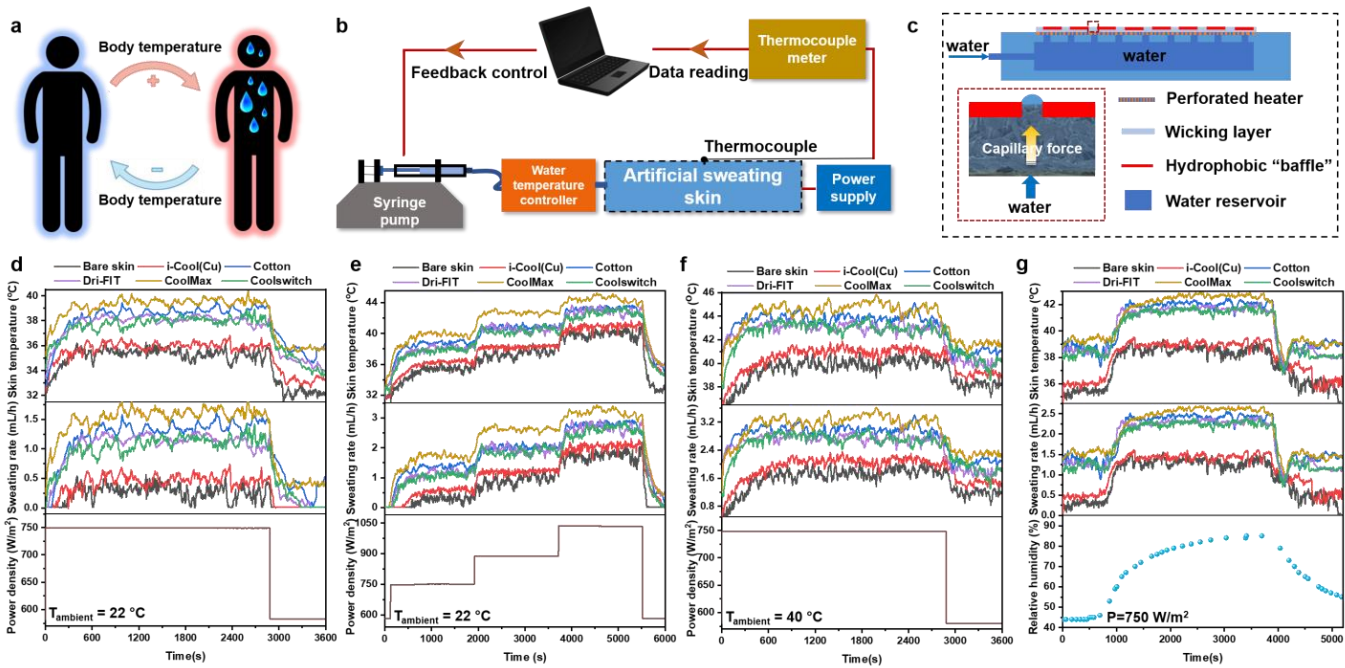

**Figure 3 | Artificial sweating skin platform with feedback control loop and measurements on it. a,** Schematic of human body temperature self-regulation mechanism. When body temperature increases, human body perspires to cool down its own temperature, which leads to reduction or suspension of perspiration in reverse. **b,** Schematic of the artificial sweating skin platform with feedback control loop simulating human body temperature self-regulation mechanism. **c,** Schematic of the detailed structure of the artificial sweating skin. The schematic in the red dash box shows the working mechanism of the modified Janus-type wicking layer which realizes uniform sweating mimicking human skin sweating scenario. **d-g,** Measurement results of skin temperature and sweating rate with i-Cool (Cu) textile and other conventional textiles under different simulated conditions: perspiration under exercise with constant skin power density and stop exercise at ambient temperature of 22 °C (**d**), perspiration under different skin power densities to simulate different exercise conditions at 22 °C ambient temperature (**e**), perspiration at high ambient temperature (40 °C) (**f**), perspiration at

varied ambient relative humidity (**g**). Ambient relative humidity in (**d**)-(**f**): 40% ± 5%. Difference of skin temperature between the i-Cool (Cu) textile and other textile samples, Welch's t-test P < 0.001, indicating strong statistical significance.

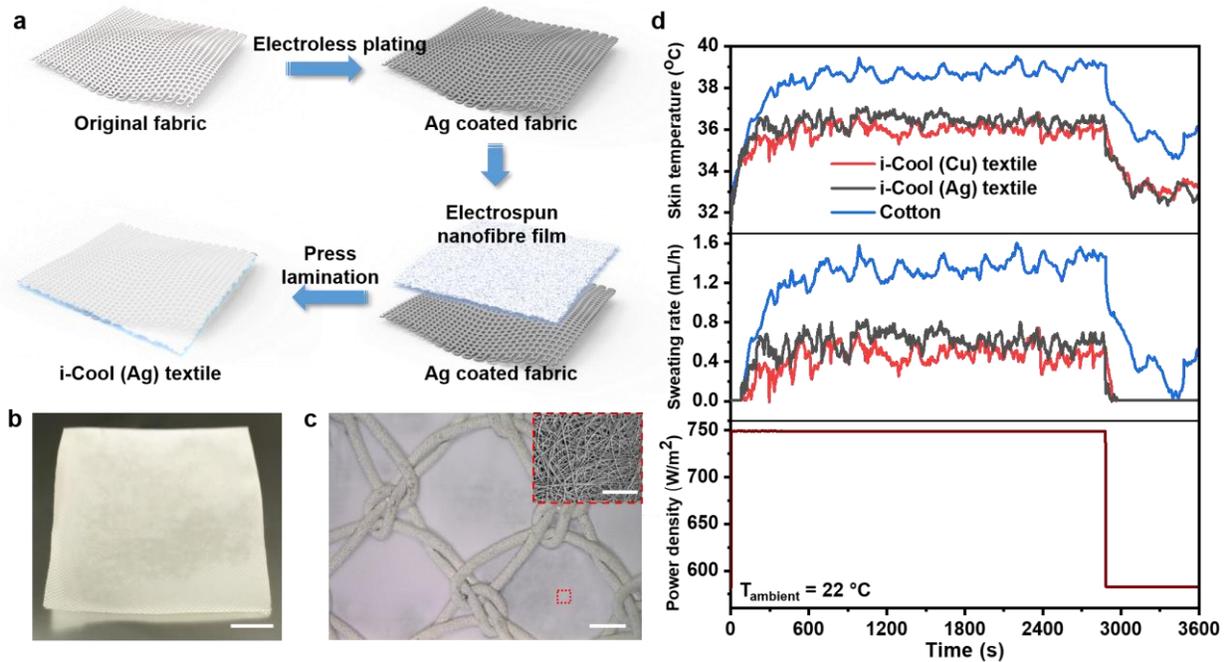

**Figure 4 | Application of the i-Cool functional structure design on practical textile. a,** Illustration of the fabrication process of i-Cool textile based on commercially available fabric. **b,** Photograph of as-fabricated i-Cool (Ag) textile. Scale bar, 1 cm. **c,** Optical microscope image showing its bottom structure. The uniform Ag coating and embedded nylon 6 nanofibres in the Ag coated PET matrix can be clearly observed. Scale bar, 200 μm. Inset shows SEM image of the nylon 6 nanofibres in heat conductive matrix pores. Scale bar, 5 μm. **d,** Measurement results of the i-Cool (Ag) textile, i-Cool (Cu) textile and cotton on the artificial sweating skin platform with feedback control loop. The similar performance of i-Cool (Ag) textile and i-Cool (Cu) textile demonstrates the feasibility of practical application of the i-Cool functional structure design.

**Supplementary Note 1. Water vapour transmission property for textiles**

Water vapour transmission, namely breathability[1,2], is an essential property for textiles for ensuring wearing comfort. The human body releases water vapour all the time, no matter whether sensible perspiration happens. At a dry state, water vapour loss from human body (insensible perspiration) accounts for about 20 percent of the total heat dissipation[3,4]. As human body heat load increases, human body starts to secrete sweat to release excessive heat. During a slight perspiration, sweat can almost be evaporated fast on the skin directly and nearly only water vapour passes through the textile. Thus, decent water vapour transmission rate of textiles are necessary for achieving good cooling effect in these situations. We characterized the water vapour transmission rate of textiles using the upright cup method (ASTM E96). As shown in Extended Data Fig. 1a, the i-Cool (Cu) textile shows comparable water vapour transmission rate to other commercial textiles such as cotton textile, which have been widely accepted as a good water vapour permeable textiles. We also performed the water vapour thermal measurement, in which sweat was evaporated directly on the simulated skin surface and only water vapour passes through textiles (Extended Data Fig. 1b). The measured skin temperature is overall negatively correlated with the water vapour transmission rate, which indicates the higher water vapour transmission rate can generally achieve better cooling effect for the skin even though water vapour condensation and textile dry thermal resistance can also influence the measured skin temperature in the water vapour thermal measurement. Moreover, measurements of evaporative resistance based on ISO11092:2014/ASTM 1868-17 were performed, and the apparatus schematic and results are exhibited in Extended Data Fig. 2. All of the tested textiles can be considered as very good in breathability according to the Hohenstein comfort rating[5]. The above test results indicate that the i-Cool (Cu) textile and other commercial textiles exhibit decent water vapour transmission ability so that they all can perform well in human

body's mild scenarios (dry state and slight perspiration). However, how textiles handle the liquid sweat and releases its evaporative cooling power should be emphasized in the moderate/profuse perspiration scenario.

**Supplementary Note 2. Thermal resistance measurement with cut bar method**

The cut bar method used in measuring the thermal resistance of these sheet-like samples is a standard technique adapted from the ASTM-5470 standard[6], using an apparatus described elsewhere[7]. As shown in Extended Data Fig. 6, the measurement principle relies on uniform steady state 1-D heat conduction to measure the heat flow and temperature drop across a sample (either a single sheet or a stack of multiple sheets). In this setup, eight thermocouples are inserted into the center of two 1×1 in.² copper reference bars to measure the temperature profiles $T(z)$ along the top and bottom bar. A resistance heater generates a heat flux $q$ which flows through the top bar followed by the sample and then the bottom bar after which the heat is dissipated into a large heat sink. The temperature profiles of the top and bottom copper bars are then used to determine both the heat flux $q$ through the sample and the temperature drop across the sample stack, $\Delta T_s = T_H - T_L$. The entire apparatus (top bar, sample, bottom bar) is wrapped in thermal insulation. A modest pressure of approximately 15 psi was applied at the top bar ($P_{applied}$) to reduce contact resistance, and no thermal grease was used due to the material porosity. At these pressures the i-Cool does not compress visibly while the other textiles do. The thermal resistances of uncompressed textiles would be even larger than those measured here and reported in Fig. 2d of the main text, making the relative performance of i-Cool (Cu) even more impressive. In operation, the temperature profile $T(z)$ of the thermocouples is measured and $T_H$ and $T_L$ are determined by extrapolating the

measured $T(z)$ profiles to the respective bars' surfaces. Also, using the known thermal conductivity of copper $k_{Cu}$ the heat flux passing through the copper bars is calculated from Fourier's law:

$$q = -k_{Cu} \frac{dT}{dz} \quad (1)$$

The $q$ calculated for upper and lower bars can differ slightly due to minor heat losses from the upper bar[7]; for this reason, we use the $q$ calculated for the lower bar for the rest of the analysis. As mentioned previously, the temperature profiles along the two copper reference bars are used to calculate the temperature drop between bar surfaces by linear extrapolation. Finally, the total thermal resistance from $T_H$ to $T_L$, including the sample(s) plus contacts, is then calculated from

$$R_{TOT} = \frac{\Delta T_s}{q} \quad (2)$$

The diagram in the red dash box in Extended Data Fig. 6 illustrates the thermal resistances for a typical stack of $N=3$ samples, where $R_s$ is the thermal resistance of a single sample sheet, $R_{c,\,ss}$ is the thermal contact resistance between adjacent samples, and $R_{c,sb}$ is the thermal contact resistance between a sample and a copper reference bar. For arbitrary $N$, by adding these resistances in series it is clear that

$$R_{TOT} = N(R_s + R_{c,\,ss}) + (2R_{c,sb} - R_{c,ss}) \quad (3)$$

Plotted on $R_{TOT}$ vs. $N$ axes, this equation takes the form of a straight line. The slope $(R_s + R_{c,\,ss})$ contains both the thermal resistance of the sample and the thermal contact resistance between samples, and the intercept $(2R_{c,sb} - R_{c,ss})$ contains both types of thermal contact resistance. Because fitting the line gives only two pieces of information (slope and intercept) but there are

three unknowns ($R_s$, $R_{c,ss}$, and $R_{c,sb}$), it is not possible to isolate $R_s$ separately from the effect of $R_{c,ss}$ in this measurement.

Typical measurement results for $R_{TOT}(N)$ for the i-Cool (Cu) samples gives $R_s + R_{c,ss} = 0.0001895$ m²K/W ± 9%. Because $R_{c,ss}$ cannot be measured independently, we instead roughly estimate its possible effect by considering a range of reference values for solid-solid interfaces at modest pressures[8], which are typically in the range $R_{c,ss}$ between ~$1\times10^{-5}$ to ~$1\times10^{-4}$ m²K/W. Subtracting this range of estimated $R_{c,ss}$ values from the measured $R_s + R_{c,ss}$ gives a bounding range on $R_s$ of between 0.0000895 m²K/W to 0.0001795 m²K/W.

To interpret the heat conduction pathways in the i-Cool (Cu), we developed a simple thermal resistor model depicted in Extended Data Fig. 7a. In this circuit, there are two types of parallel pathways. The first type of pathway is columns of pure polymer, of length ($l_1 + l_2$), thermal conductivity $k_p$, and total cross-sectional area $\phi A$, where $\phi \approx 0.32$ is the porosity of the copper foil and A is the total area of the sample. Similarly, the second type of pathway is columns of (copper + polymer), where each such column has $l_1$ of copper in series with $l_2$ of polymer, and the total cross sectional area is (1-$\phi$)A. Note that $A$ will cancel out at the end of the calculation since we are presenting all thermal resistances on an area-normalized basis (SI units: m²·K/W).

Solving the resistor network and rearranging, the thermal resistance of the sample per unit area [m²K/W] is

$$R_s = \left( \frac{k_p \phi}{l_1 + l_2} + \frac{1}{\left( \frac{l_1}{k_{Cu}(1-\phi)} + \frac{l_2}{k_p(1-\phi)} \right)} \right)^{-1} \qquad (4)$$

where $k_{Cu}$ and $k_p$ are the thermal conductivity of copper and nylon 6 respectively, $l_1$ and $l_2$ are the thickness of copper and nylon 6 nanofibres respectively, and $\phi$ is the porosity of the copper foil. The $R_s$ can be calculated by setting the parameter values: $k_{Cu}$ = 400 W/(m·K), $k_p$ = 0.28 W/(m·K) (midpoint of representative literature values)[9-11], $l_1$ = 25 μm, $l_2$ = 20 μm, and $\phi$ = 0.32. Using these values, the calculated $R_s$ is 8.7×10$^{-5}$ m$^2$·K/W. This model value is quite close to the low end (8.95×10$^{-5}$ m$^2$·K/W) of the range of $R_s$ as determined from the measurements combined with literature estimates for $R_{c,ss}$. Due to the high thermal conductivity of copper, changes to $l_1$ cause only a minor change to the overall thermal resistance, which indicates there is large flexibility for heat conductive matrix thickness selection. As shown in Extended Data Fig. 7b, increasing the Cu thickness ($l_1$) from 25 μm to 1000 μm (40 times) only increases the thermal resistance from 8.7×10$^{-5}$ m$^2$·K/W to 1.08×10$^{-4}$ m$^2$·K/W (~24% increase).

**Supplementary Note 3. Water outflow from the perforated hydrophilic sheet**

For a perforated hydrophilic sheet, water can be squeezed out from the heater surface easily, however, the requirement of uniform water outflow is challenging for such perforated membrane design. As displayed in Extended Data Fig. 11a, water cannot outflow uniformly from each pore (diameter at 200 μm) on the surface, even at an ultrahigh flow rate, not only because it is much easier for water to overcome the Laplace pressure at several pores firstly rather than at every pore, but also due to the complexity involving fluid dynamics and friction between water and device. Moreover, much larger pores (diameter at 3 mm) were still not able to facilitate the uniform water outflow (Extended Data Fig. 11b), which furthermore indicates that the design of perforated membrane as skin surface is not feasible enough to provide uniform perspiration condition.

**Supplementary Note 4. Thermal simulation for actual human body**

To model the thermal impacts of the textiles on human body, we set up a coupled heat and mass transfer equation, based on the previously reported model[12]. In the model, the human body is represented by two concentric shells: core and skin. Above the skin, the clothing forms another layer. The outside of the clothing is ambient air. The schematic of human-clothing-environment system is shown in Extended Data Fig. 16a. This model considers the coupled heat and mass transfer across body core, skin, inner and outer surface of textile, as well as the human body thermal responses such as heat generation, blood circulation, and perspiration[12]. This model approximates a scenario in which the textiles are not flooded with liquid.

The thermal model consists of both heat and mass transfer process, considering a four-node system consisting of body core, skin, inner surface of textiles, and outer surface of textiles, with their temperatures represented as $T_c$, $T_s$, $T_2$, $T_1$, respectively. In the heat transfer part, we consider the thermal balance at each node, starting from the body core:

$$\frac{m_c C_c}{A} \frac{\partial T_c}{\partial t} = M(T_s, T_c) - C_{res}(M, T_a) - E_{res}(M, T_a) - H_b(T_c, T_s) \tag{5}$$

Where $m_c$ is the mass of the body core, $A$ is the area of the body core, $C_c$ is specific heat of body core, $M(T_s, T_c)$ is the metabolic heat production, $C_{res}(M, T_a)$ and $E_{res}(M, T_a)$ are the convective and evaporative heat loss from respiration respectively, and $H_b(T_c, T_s)$ is the heat transfer from body core to skin via blood circulation.

In the skin layer:

$$d_s \rho_s C_s \frac{\partial T_s}{\partial t} = H_b(T_c, T_s) - E_{sw}(T_c, T_s) - E_{diff}(T_s, P_a) - h_{ti}(T_s - T_2) \tag{6}$$

Where $d_s$, $\rho_s$, $C_s$ are the thickness, density, and specific heat of the skin layer, respectively. The skin layer gain heat from body core through blood circulation $H_b(T_c, T_s)$. $E_{sw}(T_c, T_s)$ is heat loss by sweat evaporation, which is calculated by sweat efficiency ($\eta$) * latent heat of water vaporization ($L$) * sweating rate ($regsw$). $\eta$ means the fraction of sweat that is evaporated. In reality, $\eta$ might be dynamic during a perspiration process and varies with different textiles and ambient conditions. Here, the $\eta$ values adopted in this simulation is estimated from the overall ratio range of evaporated water divided by total injected water during the artificial sweating skin test experimental processes. Plus, to simplify the model, we ignore the evaporative cooling effect difference in/on different textiles, assume any water evaporation per unit mass (no matter on the skin or in/on the textile) can take away the same amount of heat from the skin and only include the sweat evaporation heat loss term ($E_{sw}$) on the skin layer. It is worthwhile to mention that evaporating unit mass of sweat in/on the textile takes away different amount of heat from the skin compared to that happening directly on the skin (i.e. cooling effect is different). Different textiles show different cooling effect as well. For example, evaporation on top of the i-Cool textile can take away heat from the skin easily due to the high thermal conductance, while evaporation on top of cotton (or other heat insulative textiles) shows lower cooling effect to the skin. $E_{diff}(T_s, P_a)$ is the heat loss by water vapor diffusing through the skin layer. $h_{ti}(T_s - T_2)$ is the heat transfer from the skin layer to the inner surface of the textile, where $h_{ti}$ is the combined heat transfer coefficient in the microclimate between the skin layer and inner surface of the textile. Here, the adopted $h_{ti}$ values are estimated from additional measurements of each textile on a dry hot plate like Fig. 2e of the main text, which includes microclimate/contact resistance between the hot plate and textile,

in series with the textile itself, and finally the convection+radiation resistance on the topside of the textile sample. Then the series thermal resistances of the textile (from the cut-bar measurements of Fig. 2d for the i-Cool (Cu), and calculated as d/K for the uncompressed cotton) and top-side convection plus radiation (as measured from a bare hotplate) are subtracted to get an estimate for $1/h_{ti}$. Thus, this approach to experimentally estimating $h_{ti}$ implicitly assumes that all three measurements (bare hot plate, i-Cool (Cu), and cotton) all have the same emissivity on their top surfaces. Overall, it must be noted that these estimates of $h_{ti}$ may carry uncertainties. But the uncertainties will not influence our conclusion significantly. Furthermore, $h_{ti}$ is also dependent on the thickness of microclimate, $d_m$, which is very difficult to assess accurately. Ideally, $h_{ti}$ should be measured between the skin and the fabric which will automatically account for the impact of the thickness of the microclimate on $h_{ti}$, however here we have assumed it to be same as that obtained from hot plate and cut bar measurement.

The inner surface of the textile is determined by:

$$d_2 C_v \frac{\partial T_2}{\partial t} = \lambda d_2 \frac{\partial C_{f2}}{\partial t} - K \frac{T_2 - T_1}{d} + h_{ti}(T_s - T_2) \tag{7}$$

$d_2$ is the thickness of the inner part the textile. $C_v$ is volumetric specific heat of the textile. $\lambda$ is the heat sorption of water vapor by the textile. $C_f$ is the water vapor concentration in the fibers. $K$ is the thermal conductivity of the textile.

The outer surface of the textile is determined by:

$$d_1 C_v \frac{\partial T_1}{\partial t} = \lambda d_1 \frac{\partial C_{f1}}{\partial t} + K \frac{T_2 - T_1}{d} - (C + R)(T_1 - T_a) \tag{8}$$

$d_1$ is the thickness of the outer part the textile. $C$ is the convective heat transfer coefficient from the textile to the surrounding air. $R$ is the radiative heat transfer coefficient from the textile to the surroundings. Here we assume the $R$ values for the i-Cool (Cu) and cotton are the same and adopted the value in previous research[12], considering the fact that the emissivity of the textiles can be modified by liquid water during perspiration. In the extreme case (dry textiles), the ~ 0.4 emissivity (ε) difference between i-Cool (Cu) and cotton will not influence our final conclusion (< 0.2 °C change for temperature difference between i-Cool (Cu) and cotton).

For the moisture transfer part, starting from the microclimate region between the skin and the inner surface of textile, the water vapor concentration in the microclimate $C_m$ is determined by:

$$d_m \frac{\partial C_m}{\partial t} = \frac{1}{\lambda} E_{sk}(T_s, T_c, P_a) - h_{ma}(C_m - C_{a2}) \quad (9)$$

Where $d_m$ is the thickness of the microclimate, $E_{sk}(T_s, T_c, P_a)$ is the total skin evaporation heat loss, and $h_{ma}$ is the mass transfer coefficient from the microclimate to the inner surface of textiles. We take $d_m \approx 10$ μm, representing very close contact between the textile and the skin. The water vapor concentration in the air filling the interfibre void space of the textile inner surface $C_{a2}$ can be determined by:

$$d_2 \varepsilon \frac{\partial C_{a2}}{\partial t} = -d_2(1-\varepsilon)\frac{\partial C_{f2}}{\partial t} - \frac{D_a}{\tau}\varepsilon \frac{C_{a2}-C_{a1}}{d} + h_{ma}(C_m - C_{a2}) \quad (10)$$

$\varepsilon$ is the fabric porosity, $C_{f2}$ is the water vapor concentration at the inner surface of the fabric, $D_a$ is the diffusion coefficient of water vapour. $\tau$ is the effective tortuosity of fabric. The water vapor concentration in the air filling the interfibre void space of the textile outer surface $C_{a1}$ can be determined by:

$$d_1 \varepsilon \frac{\partial C_{a1}}{\partial t} = \frac{D_a}{\tau}\varepsilon \frac{C_{a2}-C_{a1}}{d} - d_1(1-\varepsilon)\frac{\partial C_{f1}}{\partial t} - h_c(C_{a1} - C_{ab}) \quad (11)$$

$h_c$ is the mass transfer coefficient from the outer surface of textiles to the ambient air. $C_{ab}$ is the water vapor concentration in the ambient air. A more detailed information of the parameters are summarized in the table as follows. For parameters with two numerical values listed, the first value is for i-Cool (Cu) and the second value is for cotton.

| Symbol | Definition | Expression / Value | Unit |
|---|---|---|---|
| $m_c$ | mass of the body core | 70 [ref.13] | $kg$ |
| $A$ | area of the body core | 1.8 [ref.13] | $m^2$ |
| $C_c$ | specific heat of body core | 2968 [ref.14] | $J \cdot kg^{-1} \cdot C^{-1}$ |
| $M(T_s, T_c)$ | metabolic heat production | $M(T_s, T_c) = M_0 + 19.4 * Colds * Coldc$<br>$Colds = \max(T_{s0} - T_s, 0)$<br>$Coldc = \max(T_{c0} - T_c, 0)$<br>$M_0 = 400$ W·m$^{-2}$, $T_{s0} = 36.8$ °C,<br>$T_{c0} = 33.7$ °C<br>[ref.13] | $W \cdot m^{-2}$ |
| $C_{res}(M, T_a)$ | convective heat loss from respiration | $C_{res}(M, T_a) = 0.0014 * M * (34 - T_a)$<br>[ref.14] | $W \cdot m^{-2}$ |
| $E_{res}(M, T_a)$ | evaporative heat loss from respiration | $E_{res}(M, T_a) = 0.000173 * M * (5867 - P_a)$<br>[ref.14] | $W \cdot m^{-2}$ |
| $H_b(T_c, T_s)$ | heat transfer from body core to skin via blood circulation | $H_b(T_c, T_s) = (K_{min} + skbfu * rb * cb) * (T_c - T_s)$<br>[ref.13] | $W \cdot m^{-2}$ |
| $d_s$ | thickness of the skin layer | 1<br>[ref.14] | $mm$ |
| $\rho_s$ | density of the skin layer | 1000<br>[ref.14] | $kg \cdot m^{-3}$ |
| $C_s$ | specific heat of the skin layer | 3760<br>[ref.14] | $J \cdot kg^{-1} \cdot C^{-1}$ |
| $E_{sw}(T_c, T_s)$ | heat loss by sweat evaporation | $\eta * L * regsw$<br>$regsw = 170 * warmb * \exp(\frac{warms}{10.7})$, unit [g/m²·h]<br>$L = 0.68$, unit [W·h/g]<br>$\eta = 0.75$ and 0.35, based on experiments<br>[ref.13] | $W \cdot m^{-2}$ |

| Symbol | Description | Value/Equation | Units |
|---|---|---|---|
| $E_{diff}(T_s, P_a)$ | heat loss by water vapor diffusing through the skin layer | $E_{diff}(t) = 0.00305(256T(r_{s,4}, t) - 3373 - P_a)$ [ref.14] | $W \cdot m^{-2}$ |
| $P_a$ | water vapour partial pressure | $P_a = 611.21 exp\left(\left(18.67 - \dfrac{T}{234.5}\right) * \dfrac{T}{257.14 + T}\right) * RH$ [ref.15] | $Pa$ |
| $RH$ | Relative humidity | 20% | N/A |
| $h_{ti}$ | combined heat transfer coefficient in the microclimate | 362 and 83, based on experiments | $W \cdot m^{-2} \cdot K^{-1}$ |
| $d$ | thickness of the textile | 45 and 400, based on experiments | $\mu m$ |
| $d_1, d_2$ | thickness of the inner and outer part of textile | $d_1 = d_2 = \dfrac{1}{2}d$ | $\mu m$ |
| $C_v$ | volumetric specific heat of the textile | $1.75 \times 10^3$ [ref. 12] | $kJ \cdot m^{-3} \cdot K^{-1}$ |
| $\lambda$ | heat sorption of water vapor by the textile | 2522 [ref. 12] | $KJ \cdot kg^{-1}$ |
| $K$ | thermal conductivity of the textile | 0.24 (from Fig. 2d) and 0.04 [ref.16] | $W \cdot m^{-1} \cdot K^{-1}$ |
| $C$ | convective heat transfer coefficient of outer surface to air | 2.91 [ref. 12] | $W \cdot m^{-2} \cdot K^{-1}$ |
| $R$ | radiative heat transfer coefficient of outer surface to air | 5.23 [ref. 12] | $W \cdot m^{-2} \cdot K^{-1}$ |
| $d_m$ | thickness of the microclimate | 10 | $\mu m$ |
| $E_{sk}(T_s, T_c, P_a)$ | total skin evaporation heat loss | $E_{sw}(T_c, T_s) + E_{diff}(T_s, P_a)$ | $W \cdot m^{-2}$ |
| $h_{ma}$ | mass transfer coefficient from the microclimate to the inner surface of textiles | 3.5 [ref. 12] | $m \cdot s^{-1}$ |
| $C_f$ | water vapor concentration in the | $\dfrac{\partial C_f}{\partial t} = h_{cf} S_v (C_{fi} - C_a)$ [ref. 12] | $kg \cdot m^{-3}$ |
| | fibers | | |
| $\varepsilon$ | fabric porosity | 0.5 | N/A |
| $D_a$ | diffusion coefficient of water vapor in air | $0.242 \times 10^{-4}$ [ref.17] | $m^2 \cdot s^{-1}$ |
| $\tau$ | effective tortuosity of fabric | 2 | N/A |
| $h_c$ | mass transfer coefficient from the outer surface of textiles to the ambient air | 6.4 [ref. 12] | $m \cdot s^{-1}$ |
| $C_{ab}$ | water vapor concentration in the air | $0.02 * RH$ | $kg \cdot m^{-3}$ |

Combining the coupled heat and mass transfer model Eqs. (5)-(11), we can calculate the thermal impacts of different textiles on human body core temperature and skin temperature. Extended Data Fig. 16b shows the temperature evolution of body core temperature and skin temperature, during exercise, with the i-cool (Cu) textile and cotton textile, respectively.

Here, we assume the initial temperatures of $T_c$ and $T_s$ are 37 °C and 34 °C, respectively. $T_a$ is set as 22 °C. The calculation results indicate that, as steady state, the i-Cool (Cu) textile indeed has a better cooling effect as compared to cotton: the body core and skin temperatures of the case with the i-Cool (Cu) textile are approximately 35.4 °C, 38.7 °C, respectively. On the other hand, the body core and skin temperatures of the case with cotton are 37.3 °C, 40.6 °C, respectively. As compared to the body core temperature and skin temperature of the cotton case, obvious cooling effect is observed in the i-Cool (Cu) textile compared to cotton (about 2 °C temperature reduction), showing its great cooling potential in the human body.

**Supplementary Note 5. Artificial sweating skin test for cotton samples**

To eliminate the effect of mass density (or thickness) of textile samples, we tested cotton samples of different area mass density on the artificial sweating skin platform with a feedback control loop. Cotton samples of four different area mass density were prepared by stripping down nonwoven cotton sheets: 26.5 g/m$^2$, 32 g/m$^2$, 76 g/m$^2$, 132 g/m$^2$. Because the thickness of cotton is hard to accurately measure, we used area mass density to differentiate them. The i-Cool (Cu) textile sample is around 107.7 g/m$^2$. The 132 g/m$^2$ sample shows comparable skin temperature to the cotton textile used in the main text (130g/m$^2$), thus we assumed the structure difference (knitted vs. nonwoven) can be ignored. The cotton sample of 26.5 g/m$^2$ was very thin and had no reasonable visible opacity for practical use (Extended Data Figure 17a). We expanded the test range to such thin cotton sample in order to push the effect of area mass density to a limit. It turned out that even such thin cotton sample with extremely low area mass density still cannot achieve comparable evaporative cooling effect to i-Cool (Cu) textile sample (Extended Data Fig. 17b), which exhibited around 1.5 °C higher skin temperature than the i-Cool (Cu) textile. This testifies to the superiority of the i-Cool functional structure design.

**Supplementary Note 6. Parameters variation in the i-Cool (Cu) textile**

The average skin temperature of the plateau stage (high power density stage) for the i-Cool (Cu) textile sample in the previous experiment of Fig. 3d is used as the benchmark, and we performed experiments under the same condition. Firstly, we investigated the thickness influence on cooling effect. Due to the dominant thermal conductivity of Cu, changing its thickness shows slight impact on the performance of the functional structure (Extended Data Fig. 21a). The influence of thermal conductive matrix thickness change on the water transport process should be the dominant factor affecting the resultant cooling performance. Secondly, textile samples containing heat conductive matrix with different thermal conductivity but the same other parameters were studied, as exhibited in Extended Data Fig. 21b. It shows that material with thermal conductivity of the same order of Cu's can lead to comparable skin temperature to i-Cool textile (Cu). However, low thermal conductivity materials would result in high skin temperature even though the water transport structure was still remained. Furthermore, we varied the pore area ratio on the heat conductive matrix which referred to Cu matrix here to probe its effect. As shown in Extended Data Fig. 21c, it seems that the pore area ratio in a certain range brings about a negligible influence on cooling performance, but the cooling effect declines when the pore area ratio attains some extent. This is probably because of the reduction of the heat conductive component, which authenticates the structure advantage of the i-Cool textile once again. Moreover, the area mass density of nylon 6 nanofibres was altered (Extended Data Fig. 21d). The experimental results indicate that there exists a trade-off in the nylon 6 nanofibres area mass density choice. Decreasing its mass density suggests the thermal conduction in the nylon 6 nanofibres can be better, however its water transport and evaporation area may be influenced thus evaporative cooling effect can be depressed at the same time, and vice versa. Therefore, nylon 6 nanofibres with area mass density in the optimal extent is preferred in designing i-Cool textile for personal perspiration management.

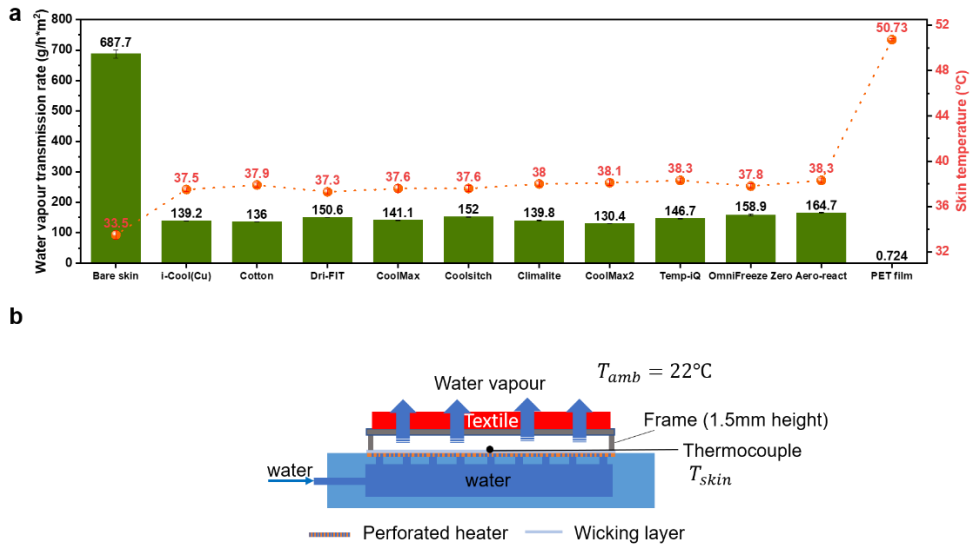

**Extended Data Fig. 1 a,** Water vapour transmission rate of textile samples, bare skin (no textile) and a PET film. The i-Cool (Cu) textile shows comparable water vapour transmission rate with other commercial textiles. Water vapour transmission rate is overall correlated with the cooling effect of the skin with different textiles in the water vapour thermal measurement. The orange dots show the measured skin temperature in the water vapour thermal measurement. **b,** Schematic of the test apparatus of the water vapour thermal measurement, in which only water vapour passes through the textile samples.

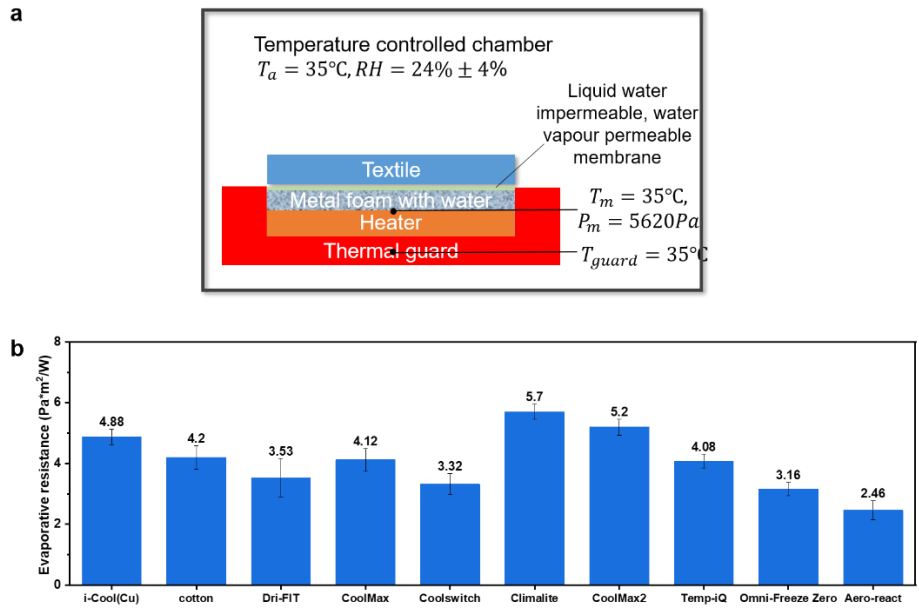

**Extended Data Fig.2 Evaporative resistance measurement of different textile samples. a,** Schematic of the test apparatus, which is according to ASTM 1868-17/ISO11092:2014 with modification. **b,** Evaporative resistance measurement results of various textile samples.

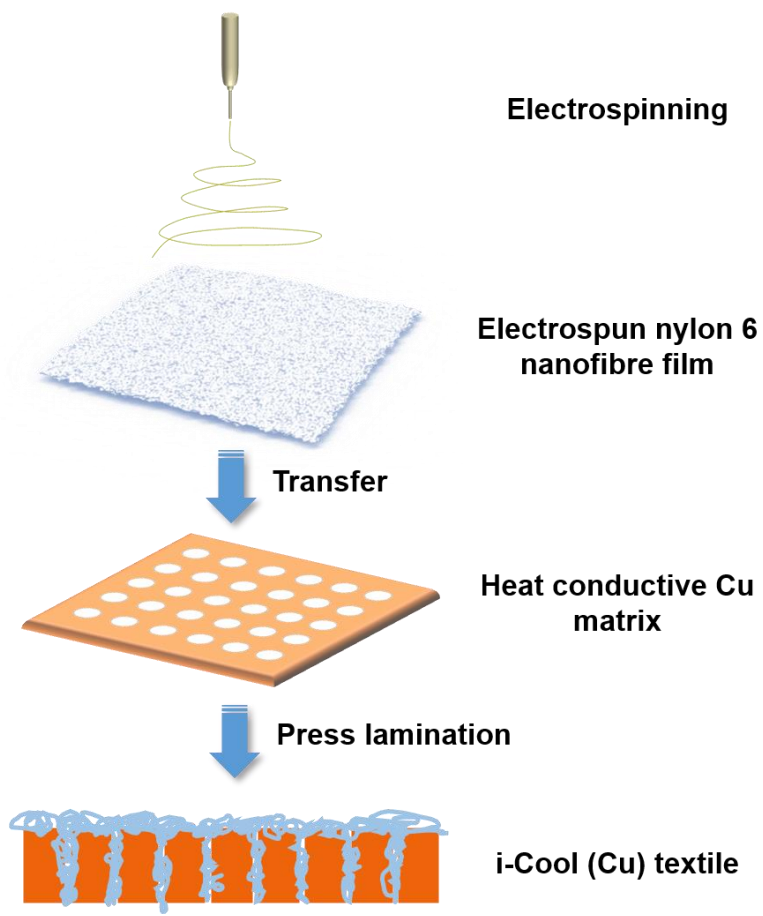

**Extended Data Figure 3.** Schematic of the fabrication process of the i-Cool (Cu) textile.

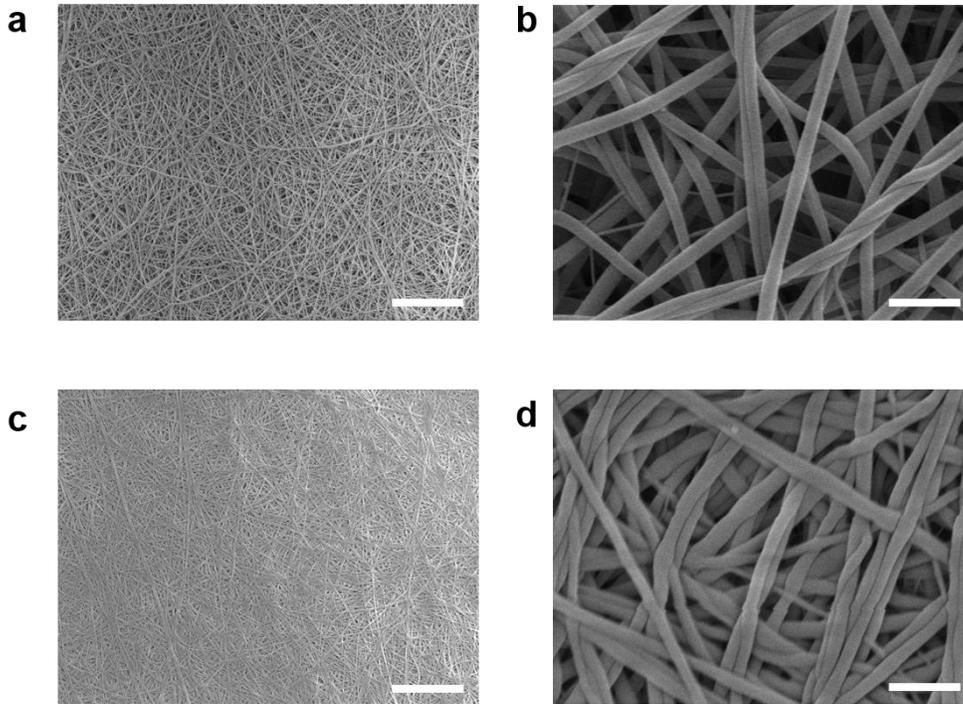

**Extended Data Figure 4. SEM images of nylon 6 nanofibres. a, b,** Nanofibres in the holes of the Cu heat conductive matrix. **c, d,** Nanofibres on the skeleton of the Cu heat conductive matrix. Scale bars in (**a**) and (**c**), 10 µm. Scale bars in (**b**) and (**d**), 1 µm.

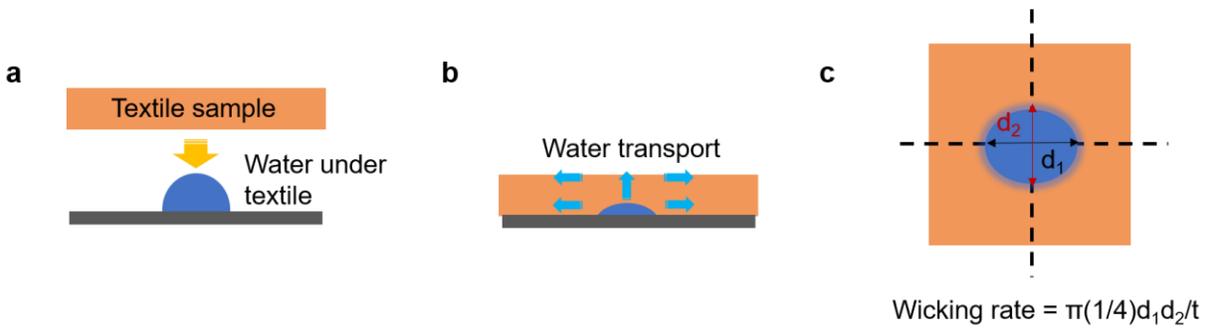

**Extended Data Figure 5. Schematic of wicking performance test method. a,** A certain amount of water was placed on a platform, and the textile sample was covered on it immediately. **b,** The water was transported in the textile in both horizonal and vertical directions. **c,** The time of water reaching a certain distance on the top surface was recorded and wicking rate was calculated using wicking area divided by wicking time.

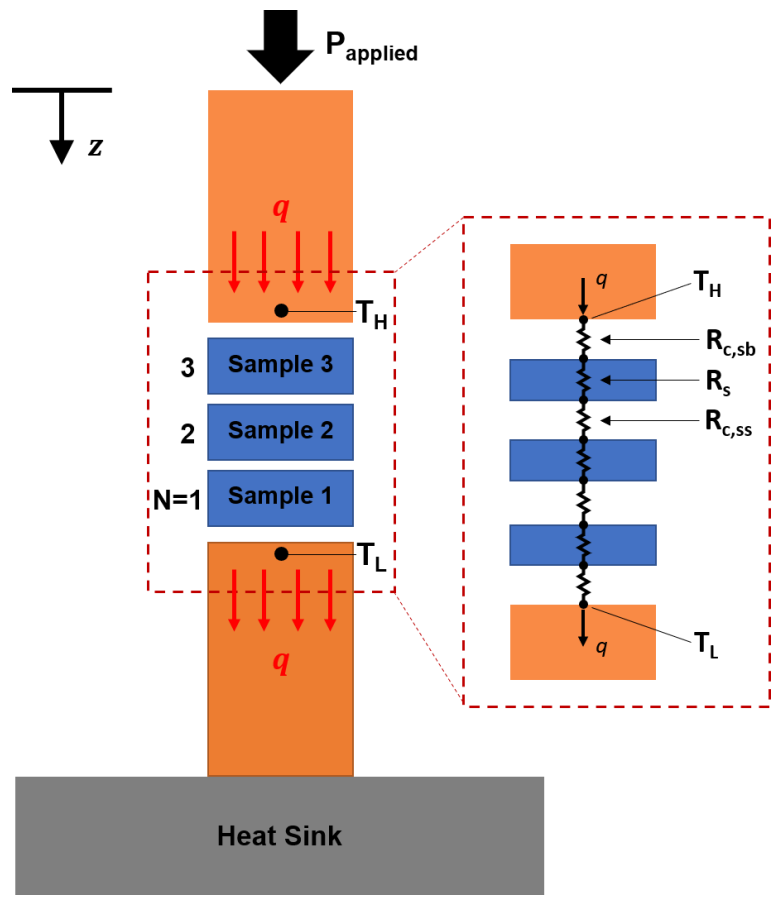

**Extended Data Figure 6. Schematic of the cut bar apparatus used in the thermal resistance measurement.** Heat flux q flows down the upper copper reference bar and through the stacked samples down through the bottom copper reference bar and is dissipated into the aluminum base heat sink. Each copper reference bar has four thermocouples inserted into the center of the bar, spaced by a pitch of 0.55 inch along the bar. The temperature readings of the thermocouples are used to determine the heat flux $q$ and the temperature drop across the samples $\Delta T_s$, which are required to calculate the total thermal resistance. The red dash box shows the three types of constituent resistances in series in the total thermal resistance measured by the cut bar apparatus when samples are stacked. $R_s$ represents the thermal resistance of a single sample sheet, $R_{c,ss}$ represents the thermal contact resistance between adjacent samples, and $R_{c,sb}$ represents the thermal contact resistance between a sample and a copper reference bar.

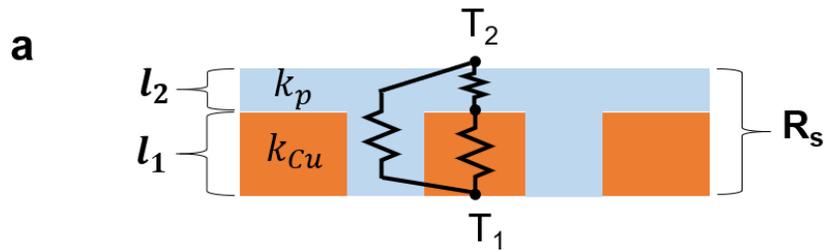

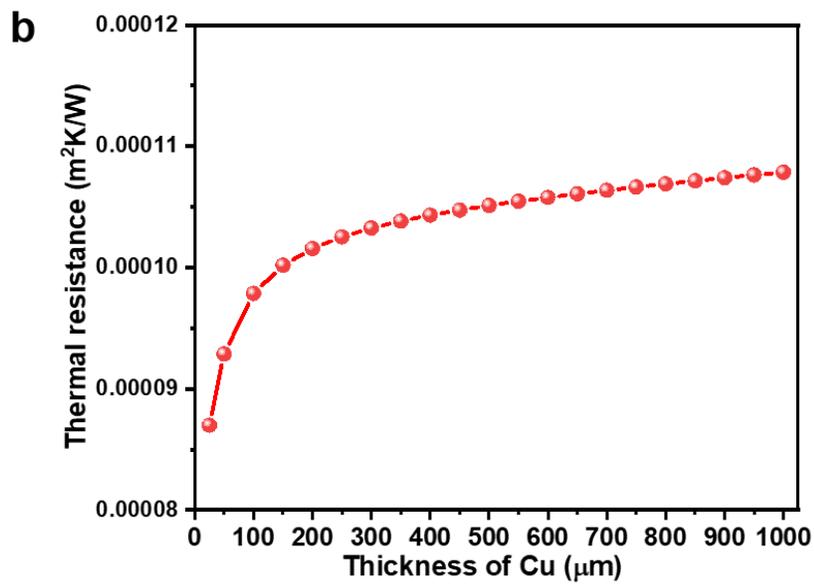

**Extended Data Figure 7. Thermal resistor model of i-Cool (Cu) textile. a,** Diagram of the series and parallel resistance network proposed in the model of the i-Cool (Cu) thermal resistance. In this schematic, the light blue section represents the nylon 6 nanofibres and the orange section represents the copper metal matrix. **b,** Simulated results of i-Cool (Cu) thermal resistance varying the thickness of Cu.

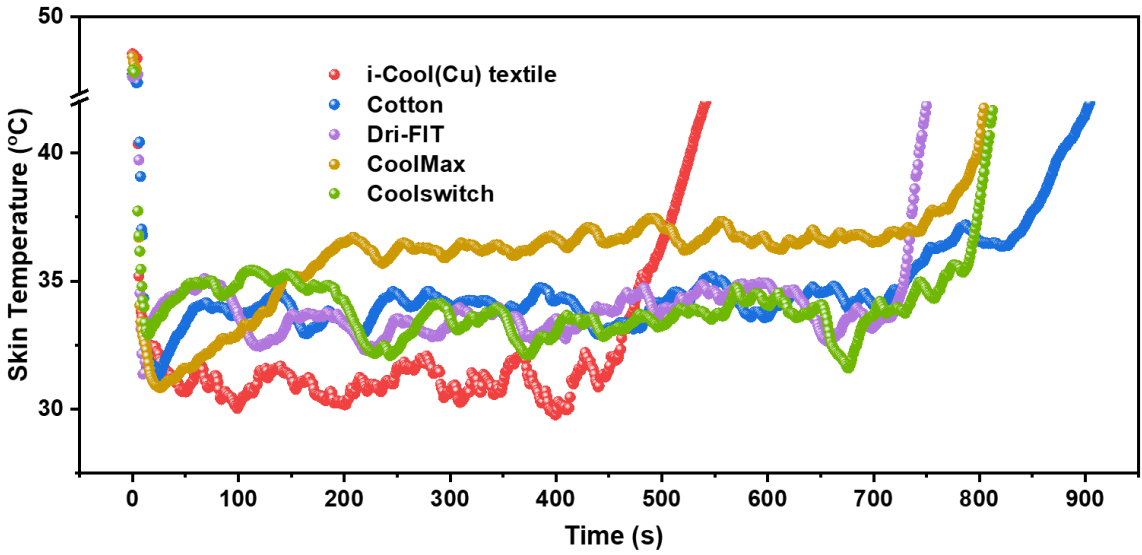

**Extended Data Figure 8.** A typical set of curves of skin temperature versus time for the i-Cool (Cu) textile and the conventional textiles during evaporation with the same initial water amount (0.1 mL) and skin power density (422.6 W/m$^2$). The i-Cool (Cu) textile shows much shorter evaporation time and lower skin temperature.

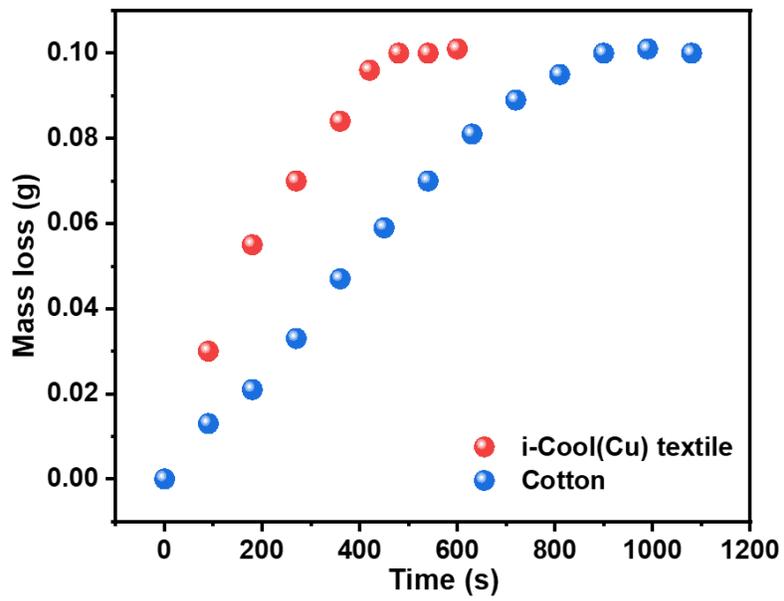

**Extended Data Figure 9.** Water mass loss versus time during evaporation with i-Cool (Cu) textile and cotton.

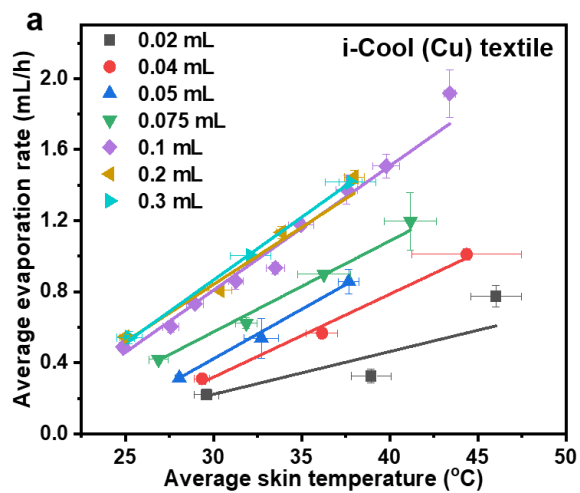 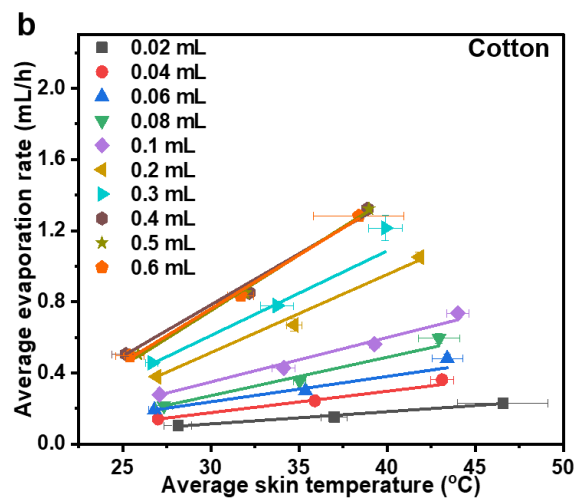

**Extended Data Figure 10.** Summarized average evaporation rate of i-Cool (Cu) textile (**a**) and cotton (**b**) with various initial water amount and average skin temperature during evaporation.

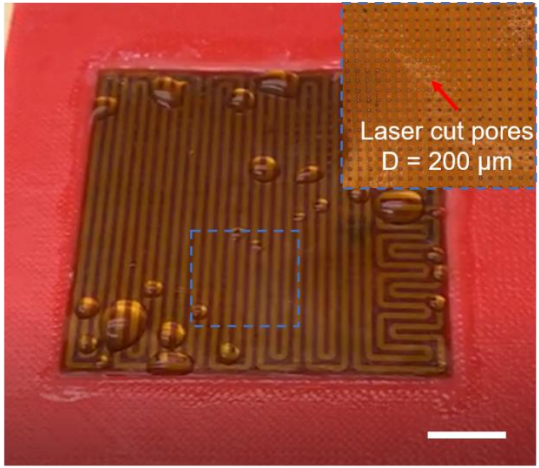 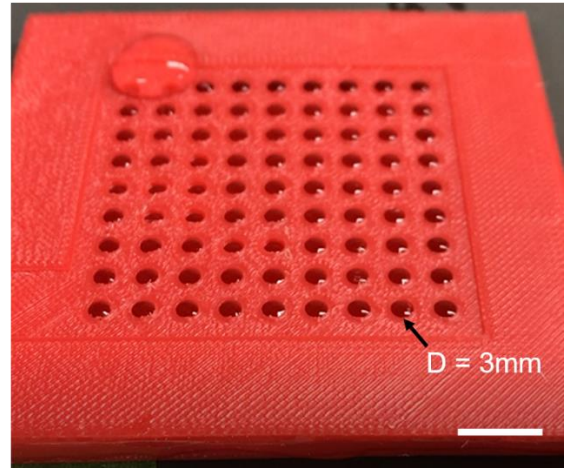

**Extended Data Figure 11.** Photographs of water outflow situation for perforated flexible polyimide heater with pores of 200 µm in diameter (**a**) and 3D printed acrylonitrile butadiene styrene (ABS) part with holes of 3 mm in diameter (**b**). For both of them, water outflow cannot be uniform. Scale bars, 1 cm. Inset of (**a**) shows the magnified photograph of the perforated heater.

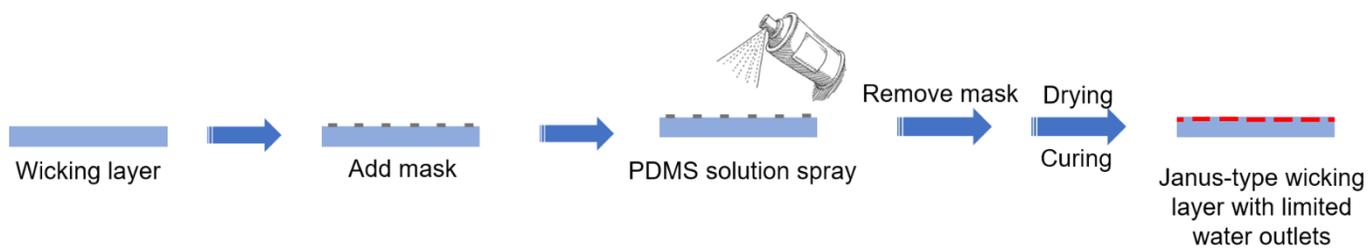

**Extended Data Figure 12.** Schematic of the fabrication process of the Janus-type wicking layer with limited water outlets.

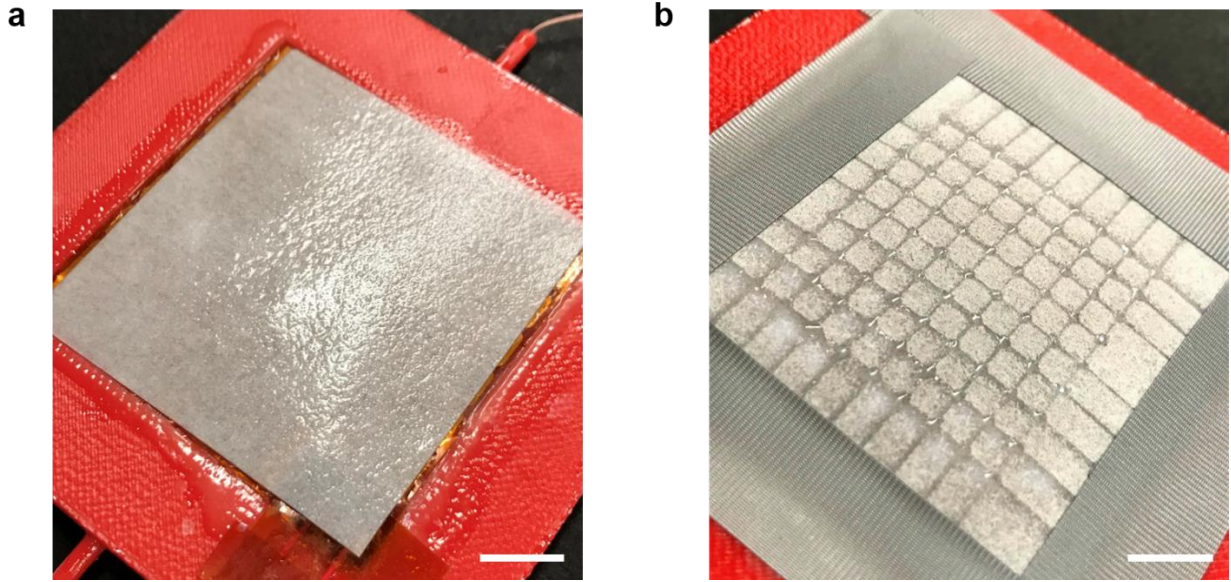

**Extended Data Figure 13. Comparison of normal wicking layer and the Janus-type wicking layer with limited water outlets. a,** Normal wicking layer can spread water totally inside itself. The surface of it is wet everywhere. **b,** The Janus-type wicking layer with limited water outlets confines water outflow into only the manufactures "sweat spots", mimicking human body sweating situation. Scale bars, 1 cm.

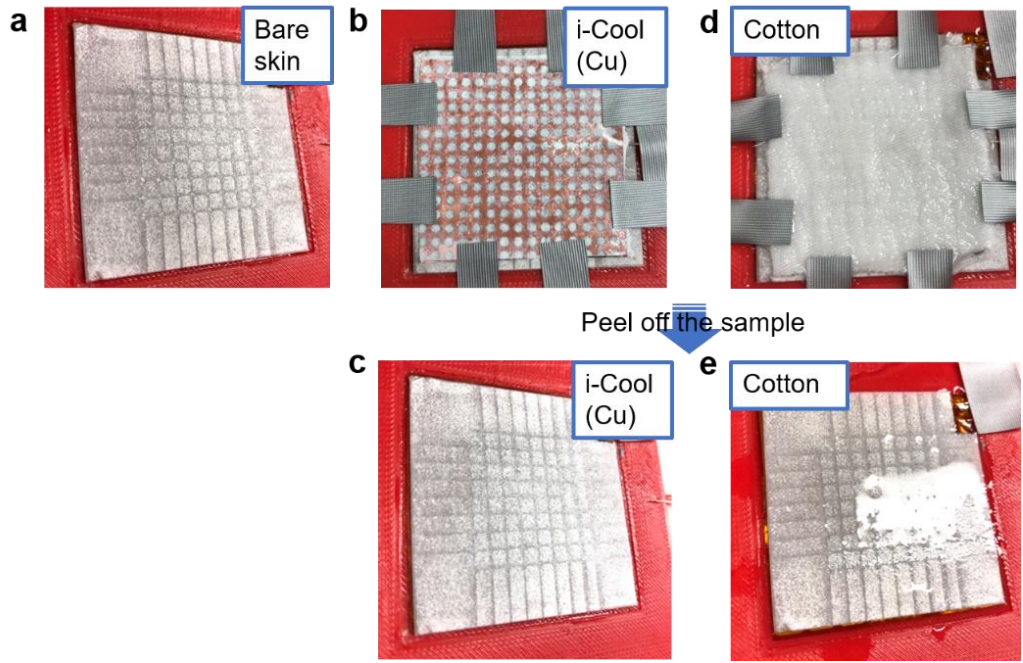

**Extended Data Figure 14.** Photographs of different samples on the sweating skin platform after a stabilization of 30 minutes, with the same skin temperature and power density, while the sweating rate for different samples was varied to keep the same skin temperature. Obviously, bare skin and the i-Cool (Cu) textile can keep skin cool using sweat much more efficiently than cotton.

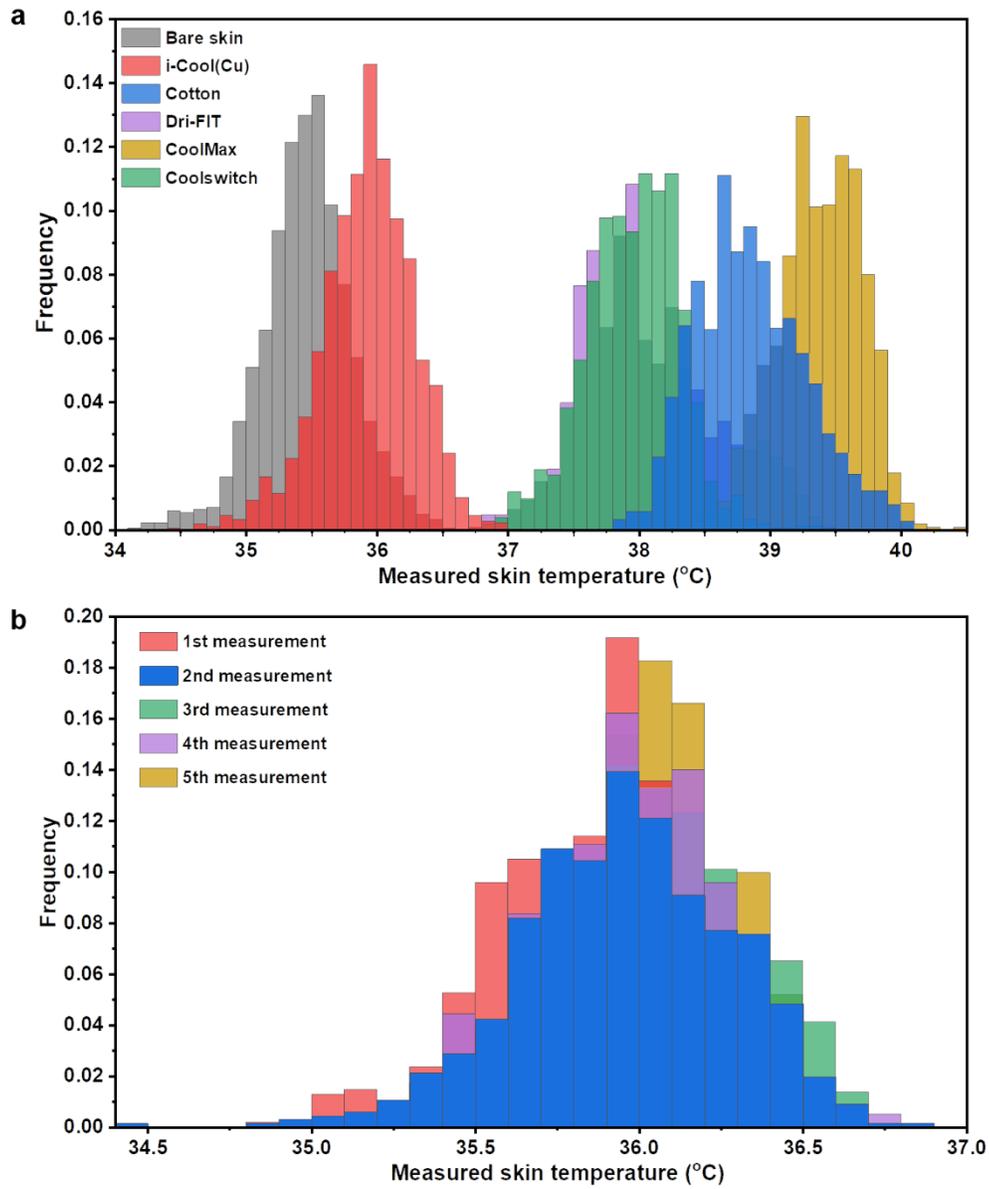

**Extended Data Figure 15.** Histograms showing data distribution in the artificial sweating skin test. **a,** Histograms of measured skin temperature with different textile samples. **b,** Histograms of measured skin temperature with the i-Cool (Cu) textile for multiple tests

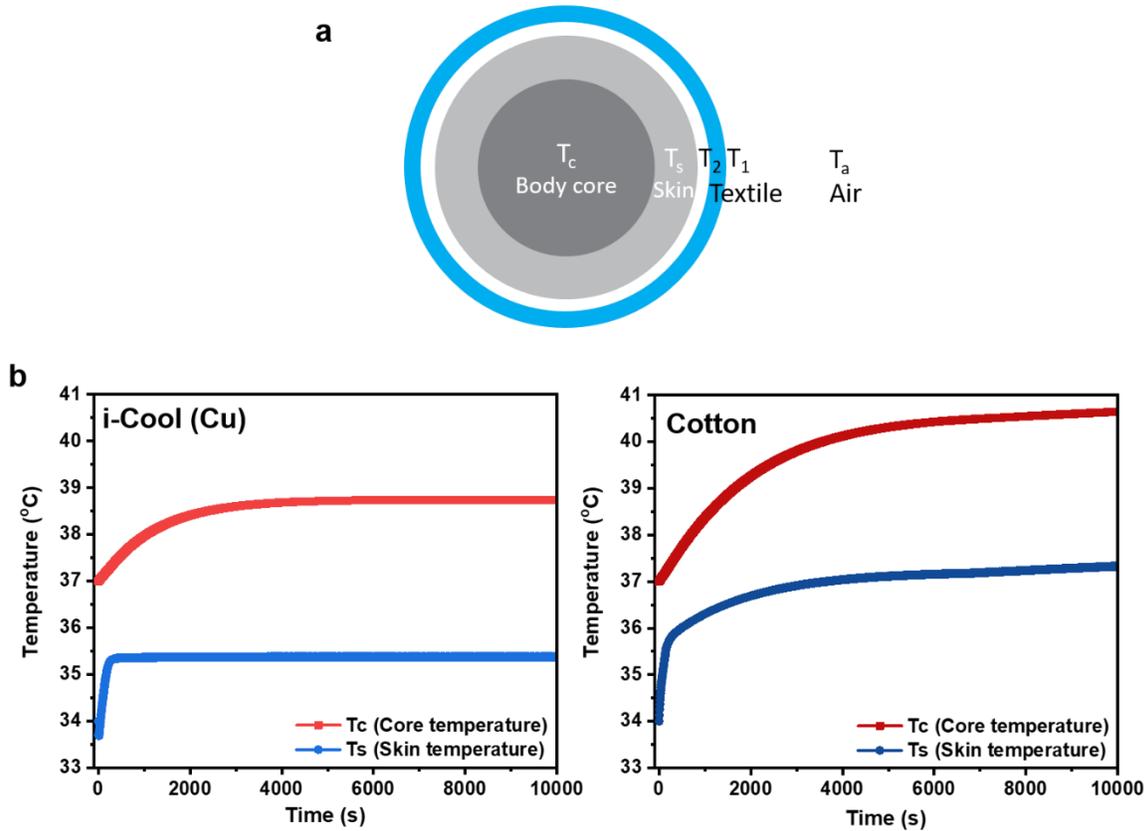

**Extended Data Figure 16. Thermal simulation for actual human body. a,** Schematics of the human-clothing-environment system. $T_c$, $T_s$, $T_2$, $T_1$, $T_a$ represent the temperatures of body core, skin, inner surface of textiles, and outer surface of textiles, respectively. **b,** Simulation results showing temperature evolution of body core temperature and skin temperature, for the cases of wearing the i-Cool (Cu) textile and cotton textile, respectively.

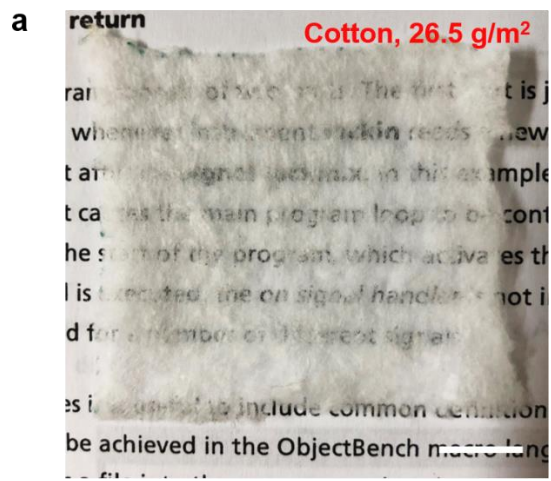
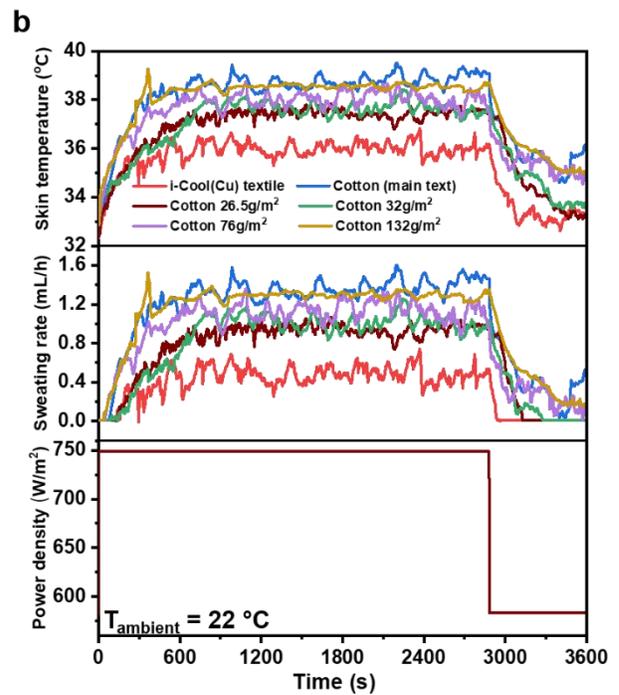

**Extended Data Figure 17. Artificial sweating skin test for cotton of various area mass density. a,** Photograph of the thin cotton sample (area mass density: 26.5 g/m$^2$). The object behind it can be clearly seen, indicating its insufficient opacity for practical use. Scale bar, 1 cm. **b,** Measurement results comparing i-Cool (Cu) textile, the cotton textile in the main text (~ 130 g/m$^2$) and cotton samples of various area mass density.

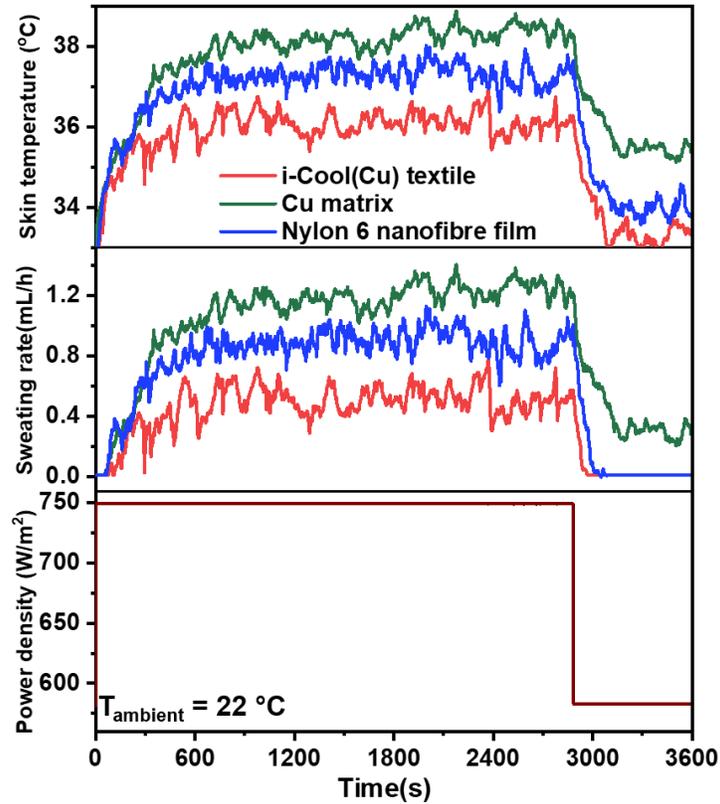

**Extended Data Figure 18.** Artificial sweating skin test for i-Cool (Cu) textile, Cu matrix and nylon 6 nanofibre film. The separate thermally conductive component and water transport component cannot show similar cooling effect to that of the i-Cool functional structure design.

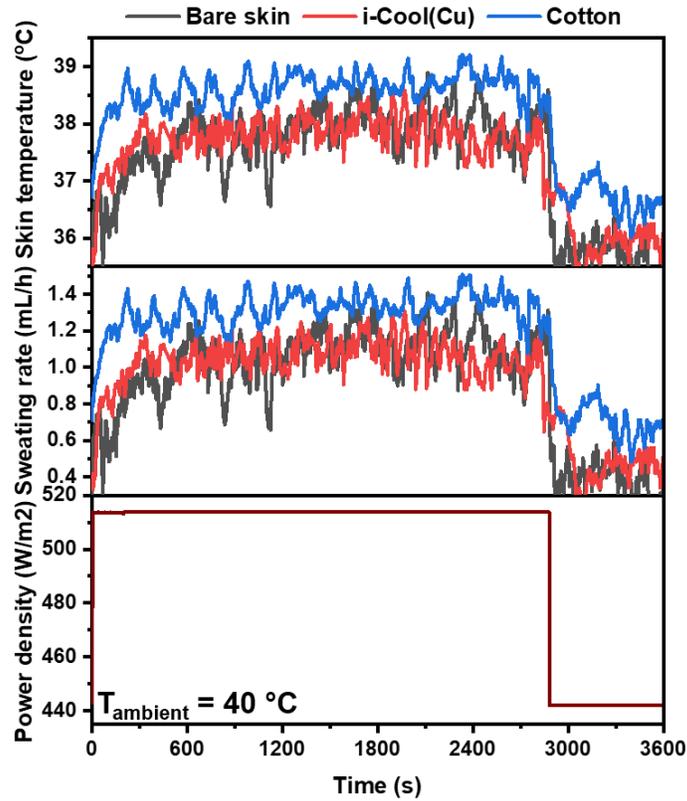

**Extended Data Figure 19.** Artificial sweating skin test for bare skin, i-Cool (Cu) textile and cotton at high ambient temperature (40 °C). The skin power density used here was adjusted to make the skin temperature for samples lower than the ambient temperature to see if the good heat conduction capability of Cu would cause a reverse effect.

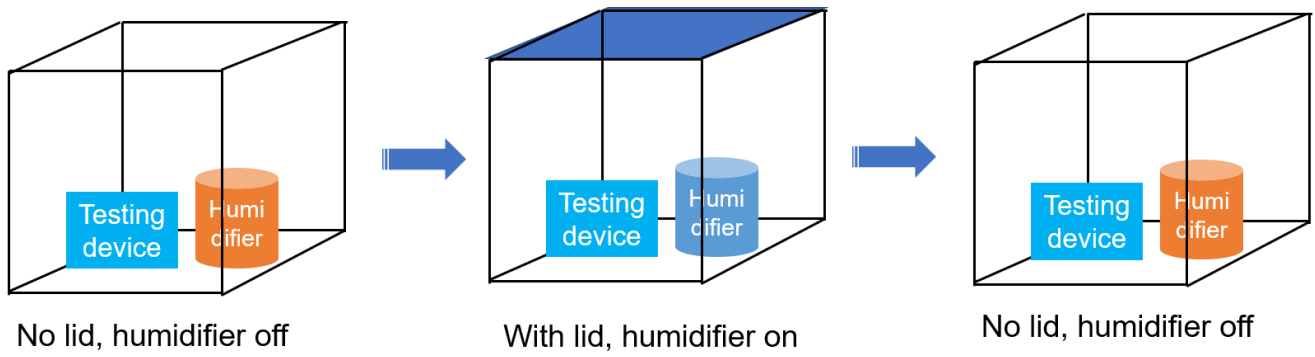

**Extended Data Figure 20.** Schematic showing the process of changing the ambient relative humidity for the artificial sweating test.

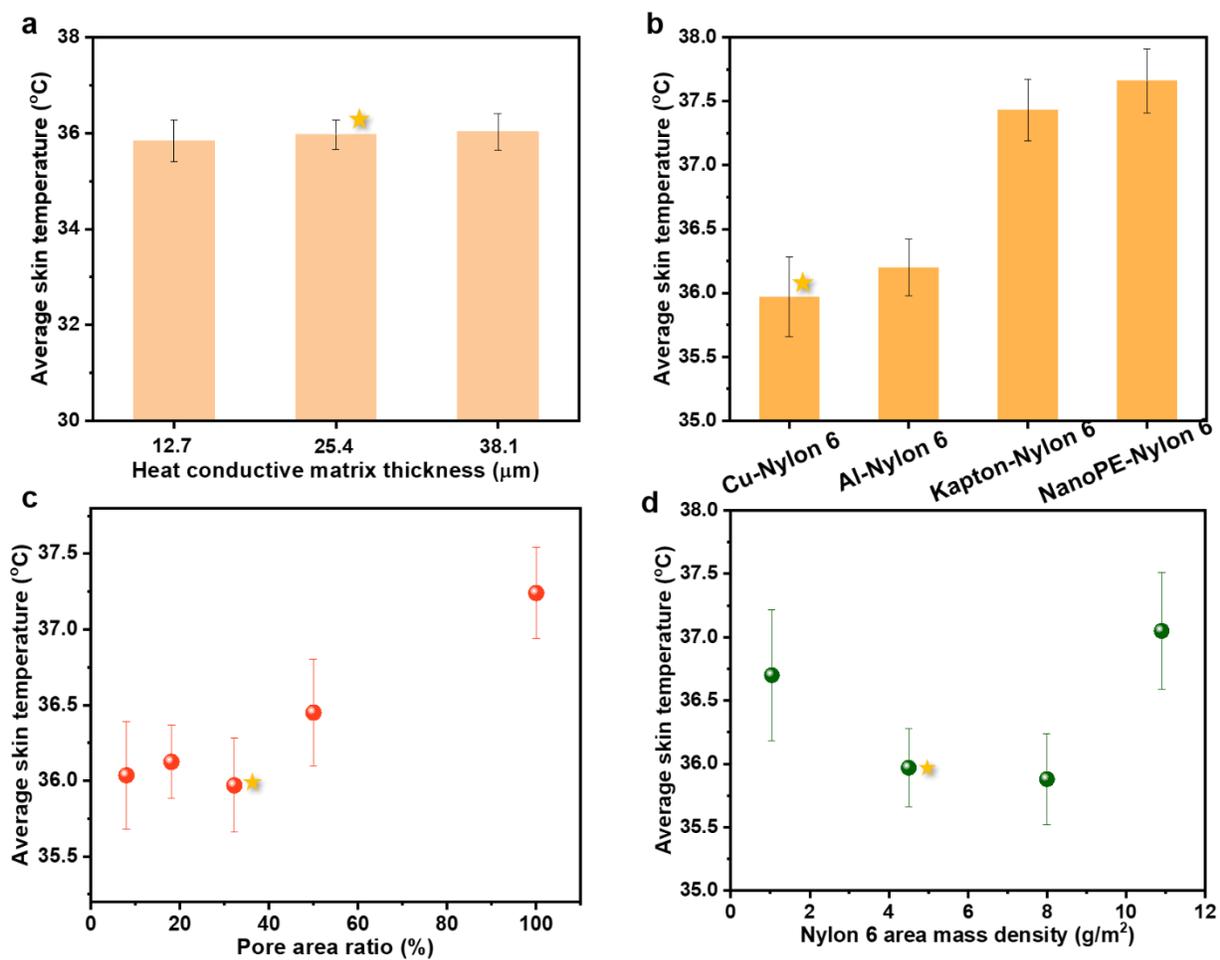

**Extended Data Figure 21. Graphs showing how structure parameters of the i-Cool (Cu) textile influence the cooling performance. a,** Experimental results of samples of three different Cu matrix thickness. **b,** Experimental results of samples with different heat conductive matrix materials. **c,** Average skin temperature for i-Cool (Cu) textile samples with different pore area ratio. **d,** Average skin temperature for i-Cool (Cu) textile samples with different nylon 6 nanofibre film area mass density. The star labels in the graphs mean the benchmark i-Cool (Cu) textile sample.

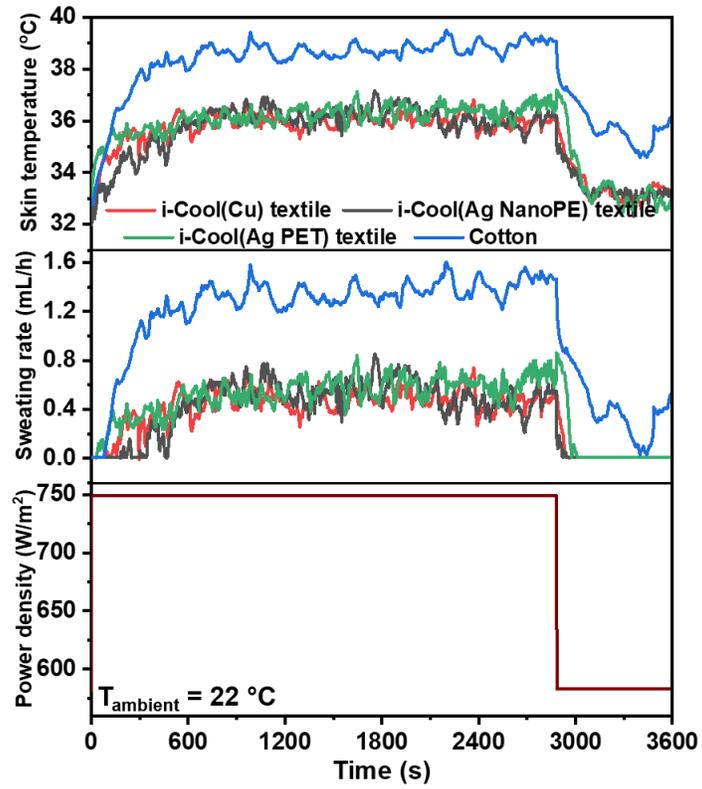

**Extended Data Figure 22.** Artificial sweating skin test for i-Cool (Cu) textile, i-Cool (Ag PET) textile and i-Cool (Ag NanoPE) textile and cotton.

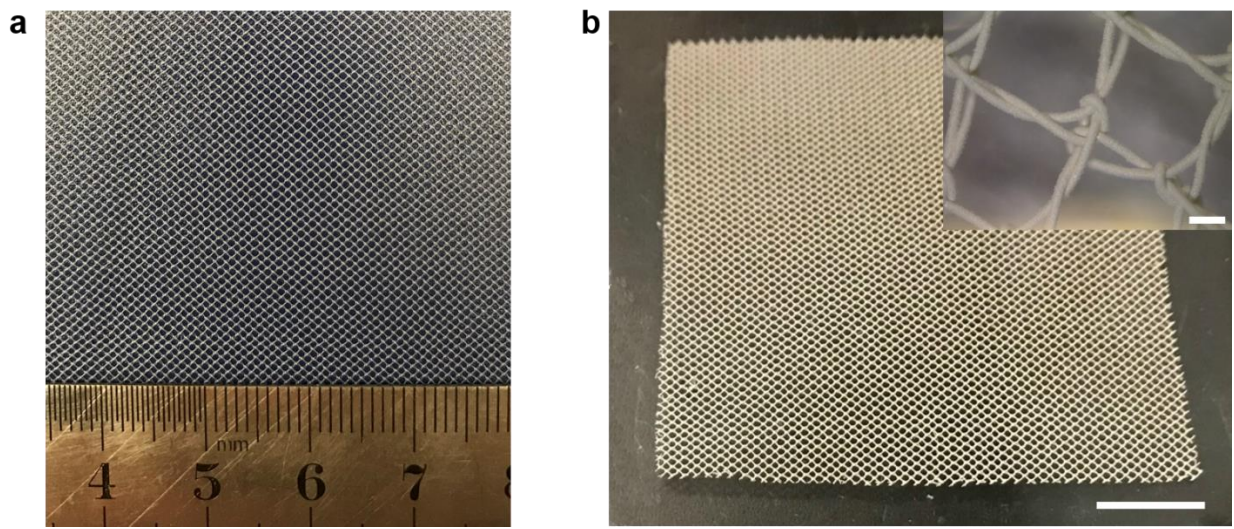

**Extended Data Figure 23. a,** The original PET fabric. **b,** The Ag coated PET fabric. Scale bar, 1 cm. The inset is the optical microscope image showing the uniform Ag coating. Scale bar, 200 µm.

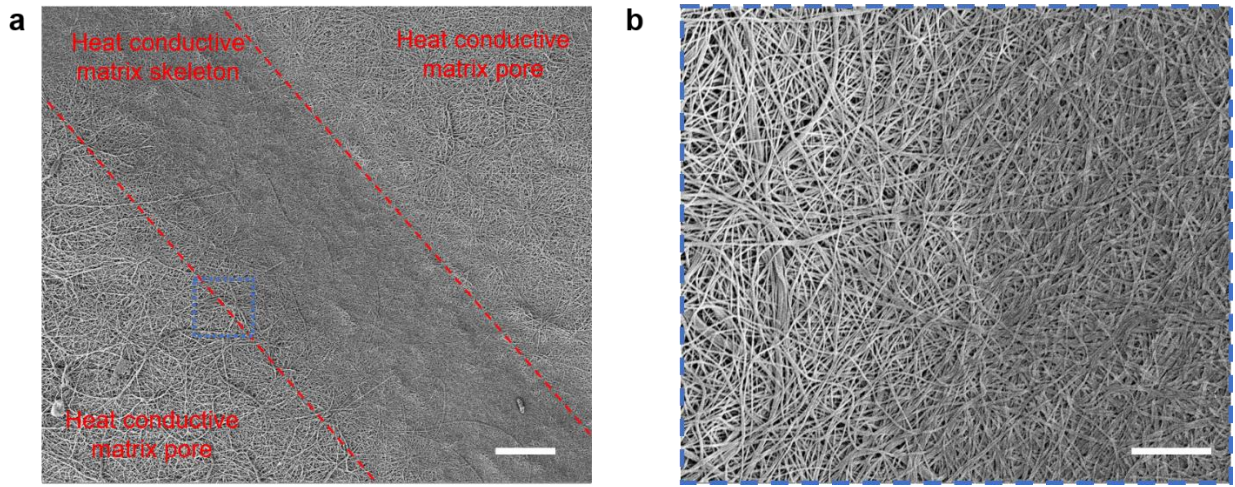

**Extended Data Figure 24. a,** SEM images of nylon 6 nanofibres in the i-Cool (Ag) textile. It is obvious to see the boundary between nanofibres on the heat conductive matrix skeleton and in the pores. Scale bar, 20 µm. **b,** Magnified SEM image of the boundary part in the blue dash box. Scale bar, 5 µm.

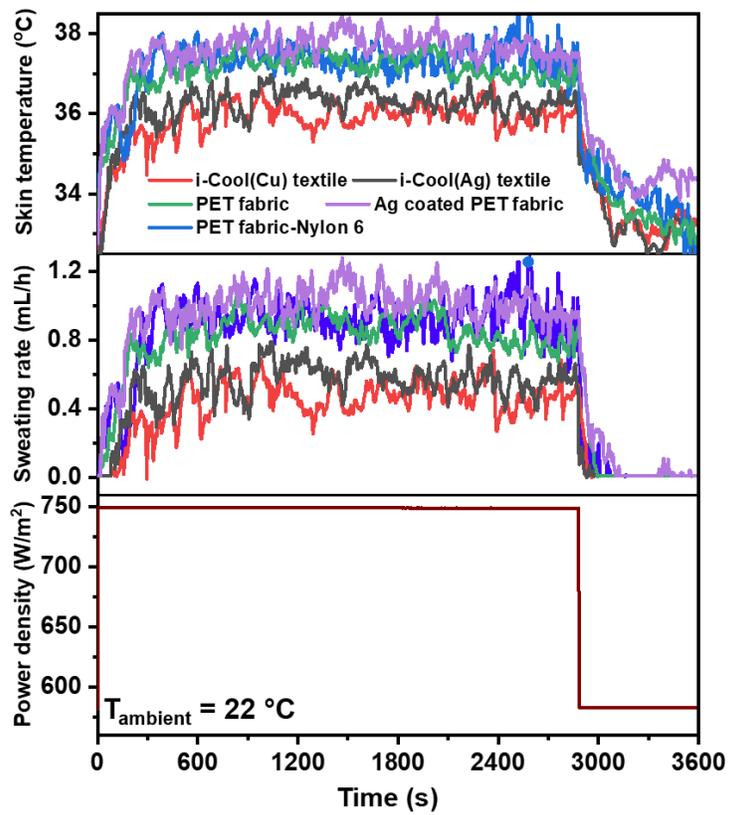

**Extended Data Figure 25.** Artificial sweating skin test for i-Cool (Cu) textile, i-Cool (Ag) textile, PET fabric, Ag coated PET fabric and PET fabric-Nylon 6.

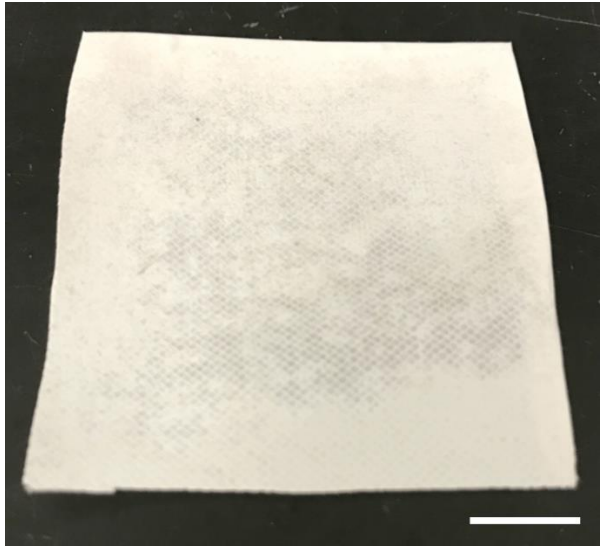

**Supplementary Figure 26.** Photograph of the i-Cool (Ag) textile after washing 50 hours. The structure remained intact. Scale bar, 1 cm.